\documentclass{article}
\usepackage[algo2e]{algorithm2e}

\usepackage[numbers]{natbib}
\setcitestyle{square}

\usepackage{microtype}
\usepackage{graphicx}
\usepackage{subfigure}
\usepackage{dblfloatfix}
\usepackage{booktabs} 
\usepackage{subfigure}
\usepackage{amssymb}
\usepackage{amsmath}
\usepackage{soul}
\usepackage{float}
\usepackage{booktabs}
\usepackage{lineno}
\usepackage{longtable}
\usepackage{graphicx}
\usepackage{setspace}
\usepackage{verbatim}
\usepackage{bbm}
\usepackage{textcomp}
\usepackage{multirow}
\usepackage[shortlabels]{enumitem}
\usepackage{mwe}
\usepackage[skip=0.5\baselineskip]{caption}
\usepackage{hyperref}
\usepackage[monochrome]{xcolor} 

\usepackage{array}
\definecolor{myblue1}{rgb}{0.1, 0.3, 0.5}
\definecolor{myred1}{rgb}{0.8, 0, 0}
\definecolor{mypink3}{cmyk}{0, 0.7808, 0.4429, 0.1412}
\definecolor{mygray}{gray}{0.6}
\newcolumntype{L}{>{\centering\arraybackslash}m{3cm}}
\newcolumntype{P}[1]{>{\centering\arraybackslash}p{#1}}
\newcolumntype{M}[1]{>{\centering\arraybackslash}m{#1}}
\AtBeginDocument{\hypersetup{ urlcolor = myblue1, anchorcolor= myblue1, linkcolor = myblue1, citecolor = myblue1}}
\usepackage[english]{babel}
\usepackage[normalem]{ulem}
\usepackage{amsthm}
\usepackage{amsfonts}
\usepackage{tabu}
\usepackage[utf8]{inputenc}
\usepackage{forest}
\newcommand{\rulesep}{\unskip\ \vrule\ }

\newcommand{\blockline}{\par\noindent\hspace{-0.1\textwidth}%
    \textcolor{white}{\rule{0.5\textwidth}{0.01pt}}\par\nobreak}

\usepackage{hyperref}

\usepackage[accepted]{icml2021}

\usepackage{calc}
\icmltitlerunning{AI-driven Hypergraph Network of Organic Chemistry: Network Statistics and Applications in Reaction Classification}

\begin{document}

\twocolumn[
\icmltitle{AI-driven Hypergraph Network of Organic Chemistry: \\ Network Statistics and Applications in Reaction Classification}

\begin{icmlauthorlist}
\icmlauthor{Vipul Mann}{co}
\icmlauthor{Venkat Venkatasubramanian}{co}
\end{icmlauthorlist}

\icmlaffiliation{co}{Department of Chemical Engineering, Columbia University, New York, USA}

\icmlcorrespondingauthor{Venkat Venkatasubramanian}{venkat@columbia.edu}

\vskip 0.3in
]

\printAffiliationsAndNotice{}  

\begin{abstract}
Rapid discovery of new reactions and molecules in recent years has been facilitated by the advances in high throughput screening, accessibility to a highly complex chemical design space, and the development of accurate molecular modeling frameworks. A holistic study of the growing chemistry literature is, therefore, required that focuses on understanding the recent trends in organic chemistry and extrapolating them to infer possible future trajectories. To this end, several network theory-based studies have been reported that use a directed graph representation of chemical reactions. Here, we perform a study based on representing chemical reactions as hypergraphs where the nodes represent the participating molecules and hyperedges represent reactions between nodes. We use a standard reactions dataset to construct a hypergraph network of organic chemistry and report its statistics such as degree distribution, average path length, assortativity or degree correlations, PageRank centrality, and graph-based clusters (or communities). We also compute each statistic for an equivalent directed graph representation of reactions to draw parallels and highlight differences between the two. To demonstrate the AI applicability of hypergraph reaction representation, we generate dense hypergraph embeddings and use them in the reaction classification problem. We conclude that the hypergraph representation is flexible, preserves reaction context, and uncovers hidden insights that are otherwise not apparent in a traditional directed graph representation of chemical reactions.
\end{abstract}

\section{Introduction}\label{sec:intro}
With the accelerated discovery of new reactions and complex molecules due to advances in computational methods, chemistry literature has been growing rapidly. The major drivers for this growth are the advances in molecule optimization, reaction engineering and optimization resulting in the discovery of novel reactions that were either unknown earlier or were infeasible, and high-throughput screening methods that have led to the re-engineering (or re-wiring) of existing reactions to make them more cost-effective and sustainable from an environmental standpoint. Hybrid AI models have a central role to play in driving chemistry growth by combining domain knowledge in the form of symbolic AI with numeric machine learning methods \cite{venkatasubramanian2022artificial}, thus leveraging the expertise of a chemist and the numeric stronghold of AI methods. Consequently, several hybrid AI-based methods have been reported for problems including thermodynamic property estimation \cite{mann2022hybrid, alshehri2021next}, reaction prediction and retrosynthesis \cite{mann2021predicting, mann2021retrosynthesis}, and chemical product design among several others as presented in the excellent review articles \cite{venkatasubramanian2019promise, zhang2020chemical, rangarajan2022towards, schwaller2022machine}.

To condense (and make sense of) the huge amount of chemistry literature that is available to us mostly in an unstructured format, we require tools that could be used to represent this knowledge in a structured format, compute coarse-grained statistics that summarize the information effectively, identify general trends on the evolution and growth of the domain, and discover new chemistry insights that were unknown earlier. While a framework that addresses these requirements could be custom-developed, network theory naturally offers tools and techniques such as -- structural statistics \cite{albert2002statistical, barabasi2003scale}, centrality measures \cite{page1999pagerank}, clustering \cite{traag2019louvain}, network embedding \cite{cui2018survey, payne2019deep}, link prediction \cite{lu2011link, maurya2021hyperedge} -- that could be used to tackle these requirements. There are several variations of graph-based representations for chemical reactions, but the most common is a directed graph representation where nodes represent molecules and directed edges from reactant nodes to product nodes represent reactions. Studies based on such dyadic representations have reported several interesting properties of the reactions network such as their scale-free network structure similar to the World Wide Web (WWW) \cite{fialkowski2005architecture}, the existence of core (most useful) and peripheral molecules across organic chemistry reactions \cite{bishop2006core}, the small-world nature of reaction networks \cite{jacob2018statistics}, which is shown to make a network robust towards node/edge deletions \cite{mann2021robust}. \cite{grzybowski2009wired, gothard2012rewiring} demonstrated applications of network theory-based studies in parallel synthesis, reactivity estimation, and rewiring of synthetic pathways.


The traditional directed graph representation for chemical reactions has several limitations. First, a directed graph representation does not capture the complete reaction context, i.e. it introduces independent directed edges for multi-reactant (or multi-product) reactions from each reactant to each product, thus losing contextual information on the presence of other reactants (or products). As a result, several seemingly independent, directed edges might correspond to the same reaction. Second, a dyadic graph representation does not allow for reaction (or edge)-specific molecular (or node) properties such as relative molecular complexity, reactivity, stoichiometry, reaction kinetics, and other properties that might be useful for making the graph representation more complete, rich, and chemistry-aware. Third, due to the above limitations, the analyses generally could not be analyzed in a self-contained manner to draw inferences and identify the trends in chemistry that are not an artifact of the reaction representation, as observed for degree correlations in \cite{jacob2018statistics}.

To address these limitations, we propose an alternative hypergraph representation where molecules are represented as vertices and an entire reaction is represented as a hyperedge. Since hypergraphs allow for an edge (or hyperedge) to connect multiple vertices together (and not just two), the entire reaction is represented using just a single, unique hyperpedge. To address the issue of incorporating reaction-specific node attributes, we use the recently proposed annotated hypergraph framework \cite{chodrow2020annotated}, which allows for each node to have hyperedge-specific annotations and makes the representation flexible to allow for reaction-specific contextual information. Therefore, compared with directed graphs, annotated hypergraphs are much more frugal in terms of the number of hyperedges, flexible in capturing reaction-level context, and due to the one-to-one correspondence between hyperedges and reactions, the statistics are self-contained, which correspond to underlying chemistry trends.

In this work, we compare and contrast the directed graph representation of chemical reactions with an annotated hypergraph representation using a standard organic chemistry reactions database containing nearly half a million reactions. Our work is the first attempt to study the network of organic chemistry using a hypergraph framework that we show to be frugal, rich, and chemistry-aware in nature, making them suitable for deriving chemistry inferences. To allow for a one-to-one comparison between the dyadic representation and the hypergraph representation, we compute standard network properties for the directed graph representation and an equivalent hypergraph representation using the same reactions dataset. At the same time, we also report the time-evolution of these properties. We also show how a hypergraph could be transformed into a weighted directed graph to allow for computation of dyadic network properties that may be ill-defined or difficult to compute for hypergraphs (at the moment). Finally, to demonstrate the use-case of such hypergraph representations not just for understanding chemistry trends but also for reaction engineering, we show how the hypergraph representation could be used in the reaction classification problem, i.e., predicting the reaction type given participating molecules which has applications in reaction mechanism generation, retrosynthetic planning, and feasibility analysis.

The rest of the paper is organized as follows -- in Section \ref{sec:intro}, we first provide a mathematical and visual description of the directed graph and hypergraph representations using an example set of four reactions in Section \ref{subsec:math_rep}, followed by a tutorial-like description of network statistics such as degree distributions, average path length, and assortativity in Sections \ref{subsec:degree-distr-motivation}, \ref{subsec:average-path-len-motivation}, and \ref{subsec:assort-motivation}; a description of the dataset used to construct the organic chemistry networks is provided in Section \ref{subsec:data_descr} and detailed network statistics along with their time-evolution and chemistry-inferences derived are provided in Sections \ref{subsec:degree-distr-results}, \ref{subsec:average-path-len-results}, and \ref{subsec:assort-results}; additional analysis based on PageRank and community detection are presented in Section \ref{sec:additional-stats-proj}. The application of hypergraph in reaction classification using reaction embeddings generated via random hyperwalks is presented in Section \ref{sec:reaction-classif-results}; finally, we present the conclusions of this study and future direction of our work in Section \ref{sec:conclusions}.

\section{Properties of directed graphs and hypergraphs} \label{sec:tutorial_properties}
In this section, we formally define directed graphs, annotated hypergraphs, and the various network statistics that we use to characterize the hypergraph network of organic chemistry. The following sections could also be treated as a tutorial that motivates various network properties using an example set of four simple reactions containing five different molecules. 

\subsection{Mathematical representation} \label{subsec:math_rep}
\begin{figure}[H]
    \centering
    \subfigure[][]{{\includegraphics[width=0.49\linewidth]{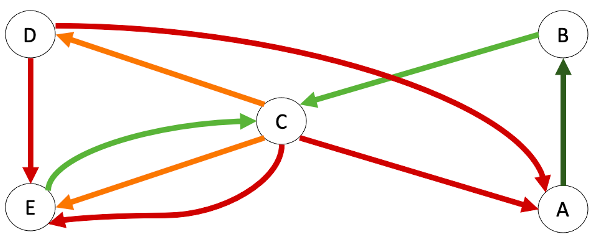}}}
    \subfigure[][]{{\includegraphics[width=0.49\linewidth]{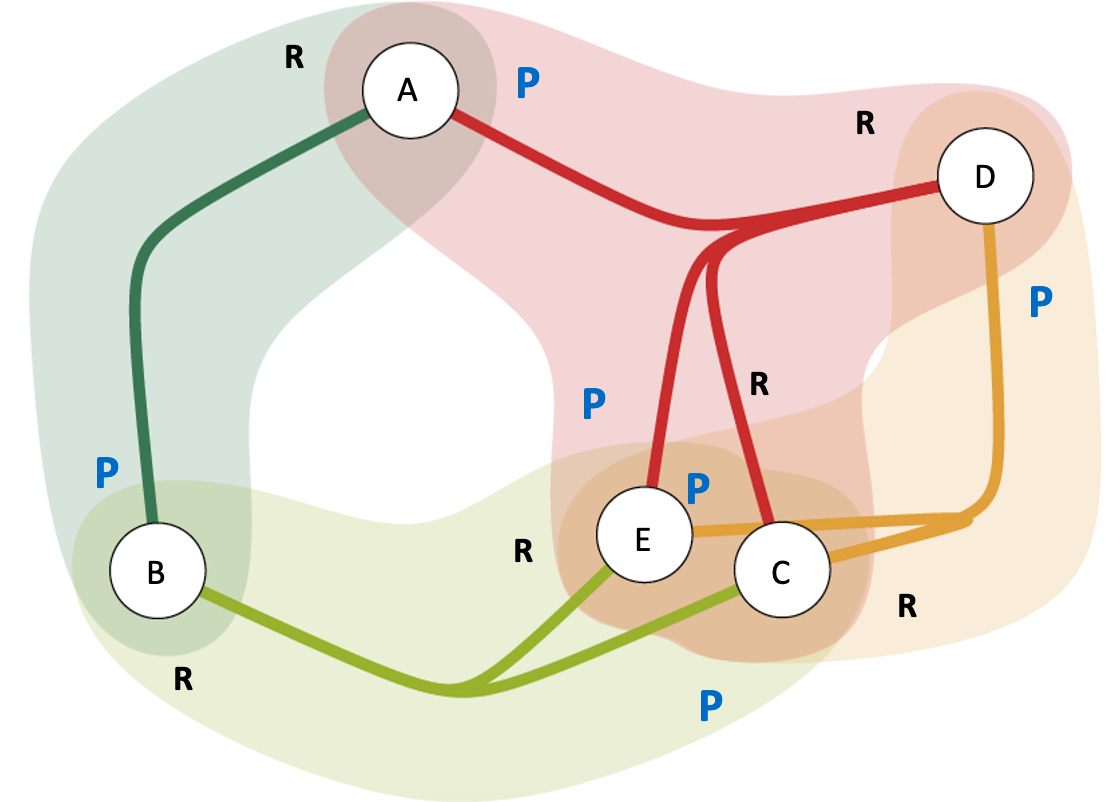}}}%
    \caption{(a) Directed graph-based representation  (b) Annotated hypergraph-based representation where an entire reaction is represented using a single hyperedge and the annotations indicate the vertex `roles' as product (P) or reactant (R)}
    \label{fig:hypergraphsViz}%
\end{figure}

A directed graph is an ordered pair $G=(V,E)$ of a set of vertices $V$ and a corresponding set of edges $E$. Each edge $e_i$ in $E$ connects a source node $s_i$ to a target node $t_i$, giving directionality to the set of edges, thus resulting in a \textit{directed} graph as opposed to an undirected graph. Chemical reactions could also be represented using such directed graphs where the reactants and products are represented as vertices, and directed edges from reactants to products representing reactions. For reactions with multiple reactants and products, the directed graph is typically constructed using all-to-all wiring with all reactants of a given reaction connecting individually to all products in the reaction through independent directed edges. Figure \ref{fig:hypergraphsViz}(a) shows a directed graph representation for the set of four reactions ($R1, R2, R3, R4$) with 5 different molecules ($A,B,C,D,E$) shown in Equation \ref{eq:example_reactions}.
\begin{equation}\label{eq:example_reactions}
    \begin{aligned}
    R1:& \quad A \longrightarrow B \\
    R2:& \quad B + E \longrightarrow C
\end{aligned}
\qquad
\begin{aligned}
    R3:& \quad C \longrightarrow E + D \\
    R4:& \quad D + C \longrightarrow A + E
\end{aligned}
\end{equation}
On the other hand, a hypergraph is a generalization of a graph where each edge is not limited to connecting just two vertices but could connect any number of vertices via hyperedges. Mathematically, a hypergraph is a pair $H=(V,E)$ where $V$ is a set of vertices and $E$ is the set of edges (or \textit{hyperedges}) where each edge contains a non-empty subset of $V$. Since each chemical reaction has contextual information about molecules along with an inherent directionality, we use \textit{annotated} hypergraphs \cite{chodrow2020annotated} with hyperedge-specific annotations (or roles) for nodes in a hyperedge. An annotated hypergraph is defined as $A=(V,E,X,l)$ where $V$ is the set of nodes, $E$ is a labeled hyperedge set where each hyperedge is a subset of $V$, $X$ is a finite label set containing the possible set of labels (or annotations/roles), and $l$ is a role labeling function for assigning roles to each edge in the label. It should be noted that each node $v$ would have a given role $x$ in given edge $e$, written as $l(v,e) = x$. Roles are contextual and they are assigned to node-edge pairs, unlike node attributes that are defined a priori for each node in dyadic graphs. For a set of chemical reactions, the set of vertices would be nodes, reactions containing the set of vertices participating in the reaction are represented as hyperedges, and the node-edge pair role could either be `product (P)'  or `reactant (R)' for nodes that play the role of reactants or products in a reaction, respectively. Figure \ref{fig:hypergraphsViz}(b) shows the equivalent hypergraph representation for the set of four reactions in Equation \ref{eq:example_reactions}.

\textit{Remark 1:} Observe that the number of (hyper)edges in a hypergraph representation is the same as the number of reactions, but this is not the case with edges in a directed graph representation.

\textit{Remark 2:} One of the primary benefits of using annotated hypergraphs is the incorporation of contextual information about reactions and molecules through hypergraph annotations or roles.

\subsection{Degree distributions} \label{subsec:degree-distr-motivation}
Degree distributions provide a general sense of the network structure and its connectivity pattern. Generating a degree distribution involves computing the degree (or number of edges) for each node and estimating the underlying probabilistic distribution that they follow. For a directed graph, each node has two kinds of degrees -- incoming degree (number of incoming edges, $d_{in}$) and outgoing degree (number of outgoing edges, $d_{out}$). The sum of the incoming and outgoing degrees, total degree ($d_{in}+d_{out}=d_{total}$), is the same as the degree of an equivalent undirected graph with directionality removed from directed edges. 

For an annotated hypergraph, equivalent degree distributions could be defined. The incoming degree for a node in the annotated hypergraph would involve counting the number of hyperedges where the node participates with a role `product' ($d_{product}$ or $d_{in}$) since products have incoming edges, and the outgoing degree would involve counting the number of hyperedges where the node participates with a role `reactant' ($d_{reactant}$ or $d_{out}$) since reactants have outgoing edges. The sum of the incoming and outgoing degrees would be the total degree ($d_{product}+d_{reactant}=d_{total}$). 

Table \ref{tab:degree_distr} shows the incoming and outgoing degrees for each node in the set of reactions in Equation \ref{eq:example_reactions} for directed graph and hypergraph representations.
\begin{table}[h]
\small
\centering
\caption{Degree distributions for the example set of reactions in Equation \ref{eq:example_reactions}}
\label{tab:degree_distr}
\begin{tabular}{@{}c|cc|cc@{}}
\toprule
\multirow{2}{*}{Node} & \multicolumn{2}{c|}{Directed graph} & \multicolumn{2}{c}{Annotated hypergraph} \\ \cmidrule(l){2-5} 
                      & \multicolumn{1}{c|}{in}    & out    & \multicolumn{1}{c|}{in (P)}   & out (R)  \\ \midrule
A & 2 & 1 & 1 & 1 \\
B & 1 & 1 & 1 & 1 \\
C & 2 & 4 & 1 & 2 \\
D & 1 & 2 & 1 & 1 \\
E & 3 & 1 & 2 & 1 \\ \bottomrule
\end{tabular}
\end{table}
\begin{figure*}[b]
    \centering
    \subfigure[][Hypergraph (H)]{{\includegraphics[width=0.22\linewidth]{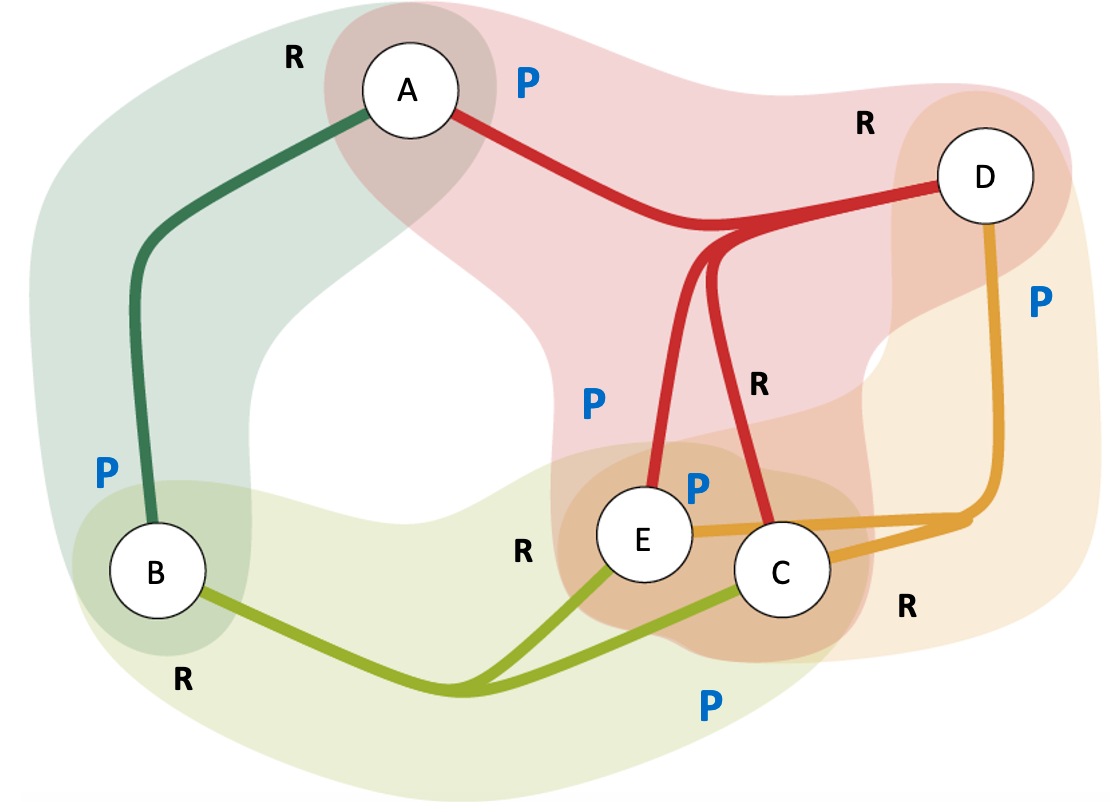}}}
    \rulesep
    \subfigure[][1-linegraph (H)]{{\includegraphics[width=0.20\linewidth]{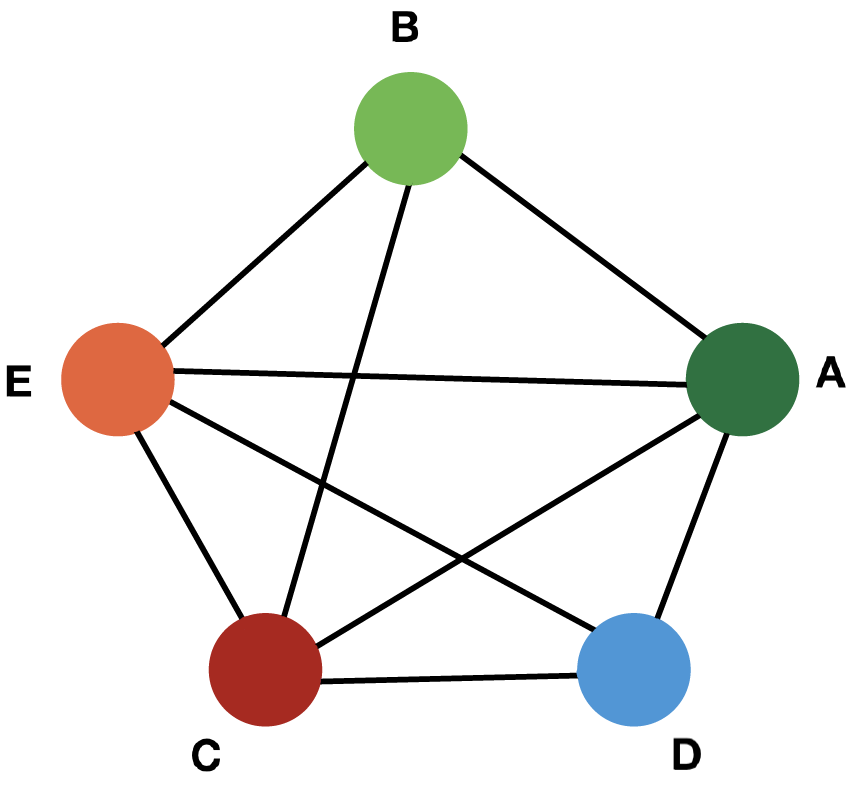}}}
    \rulesep
    \subfigure[][2-linegraph (H)]{{\includegraphics[width=0.20\linewidth]{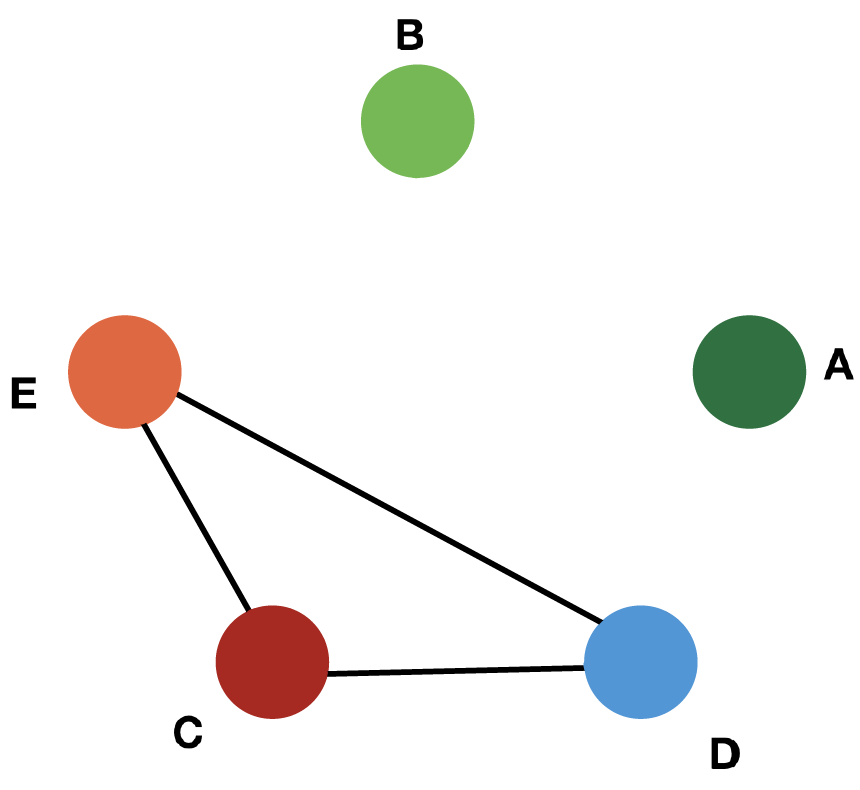}}}%
    \rulesep
    \subfigure[][3-linegraph (H)]{{\includegraphics[width=0.20\linewidth]{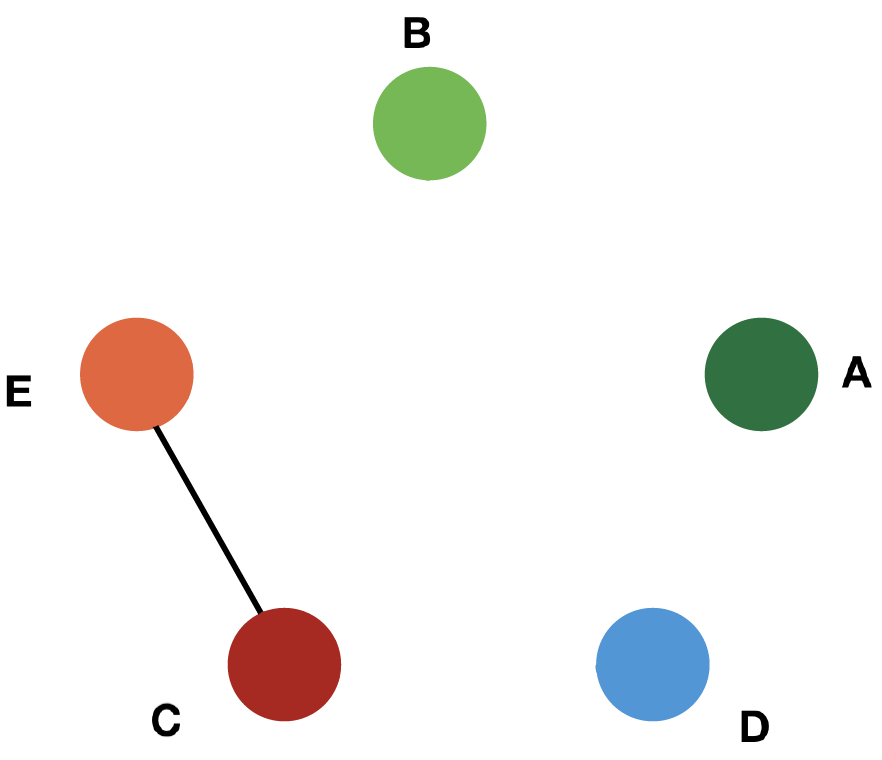}}}\\ \hrule
    \subfigure[][Dual hypergraph (H$^*$)]{{\includegraphics[width=0.22\linewidth]{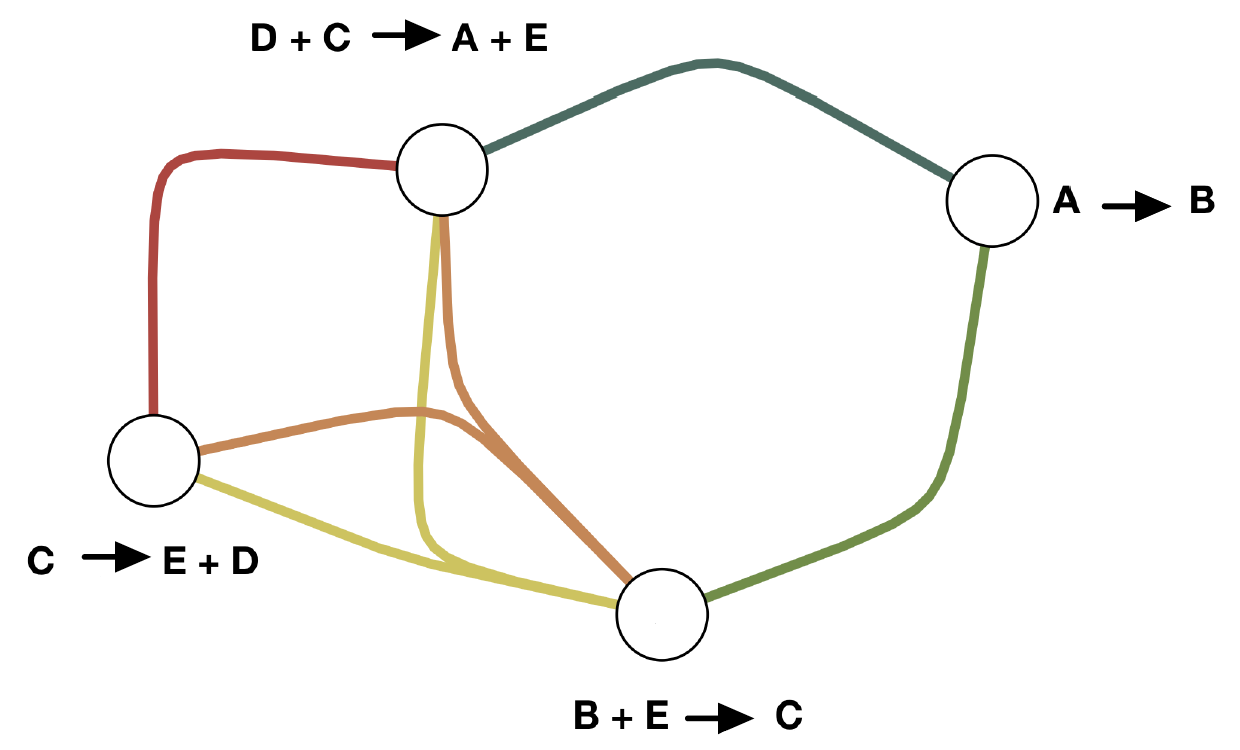}}}
    \rulesep
    \subfigure[][1-linegraph ($H^*$)]{{\includegraphics[width=0.20\linewidth]{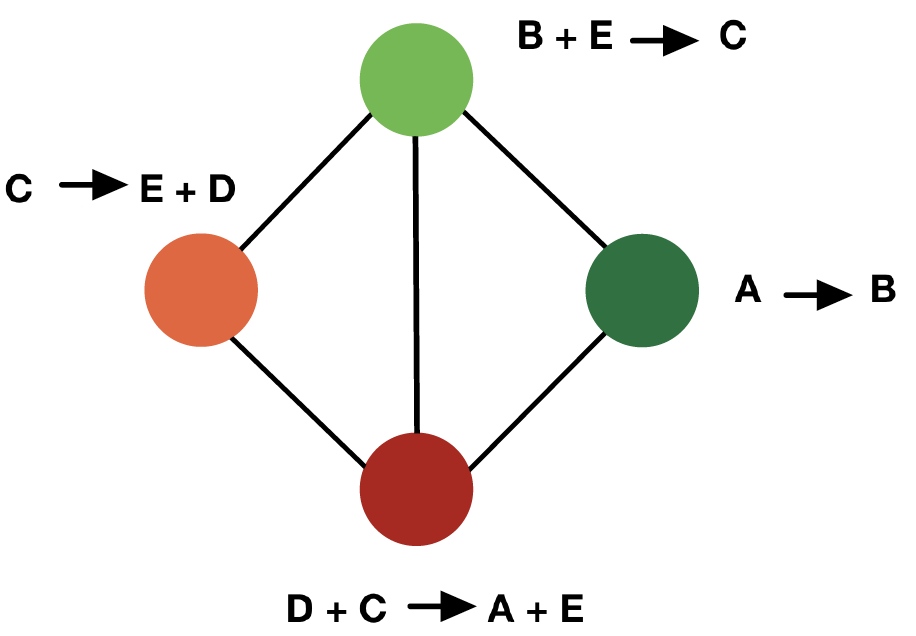}}}
    \rulesep
    \subfigure[][2-linegraph ($H^*$)]{{\includegraphics[width=0.20\linewidth]{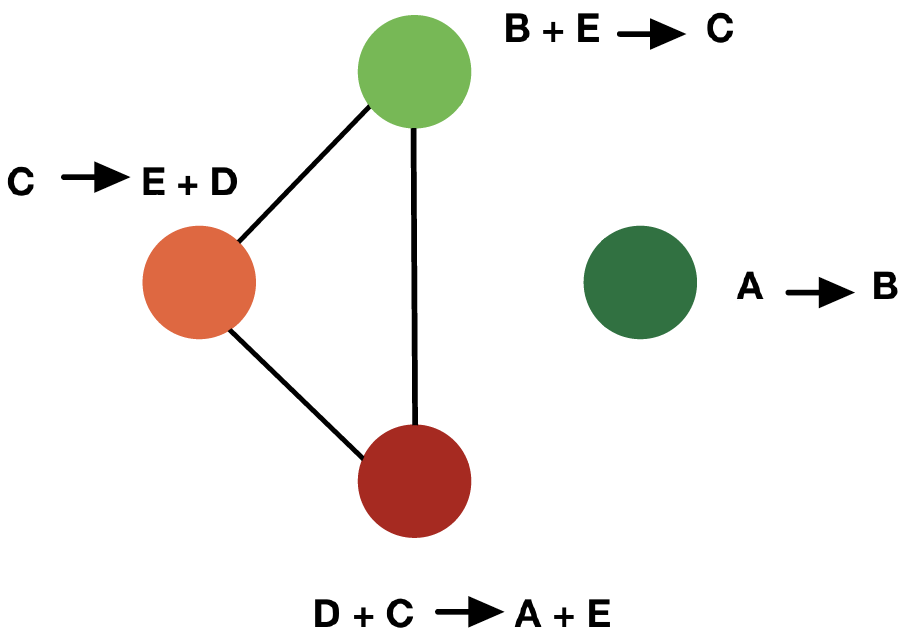}}}
    \rulesep
    \subfigure[][3-linegraph ($H^*$)]{{\includegraphics[width=0.20\linewidth]{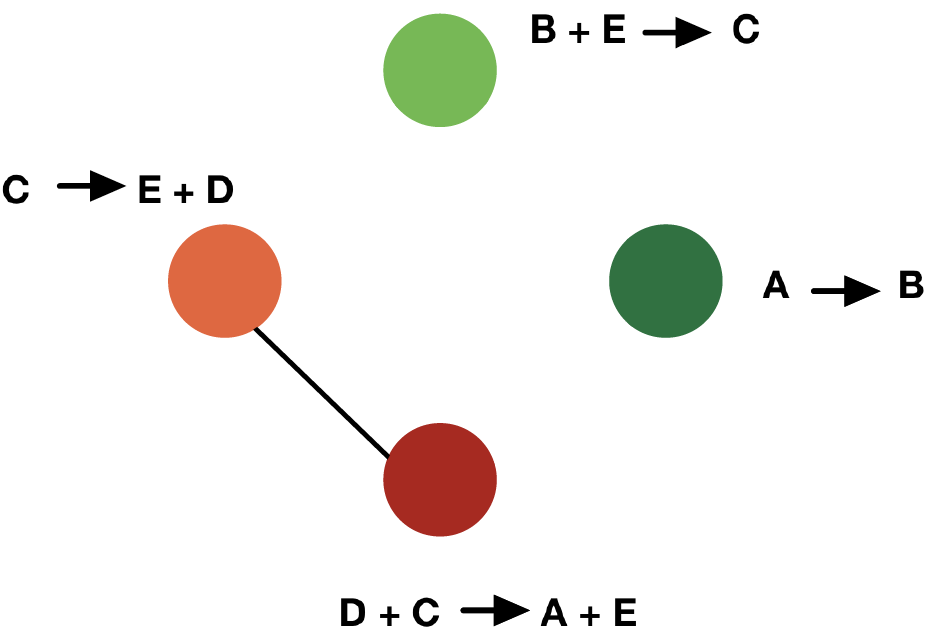}}}
    \caption{The hypergraph (H), dual hypergraph ($H^*$), and their respective s-linegraphs for the example set of four reactions in Equation \ref{eq:example_reactions}}
    \label{fig:slinegraphs}%
\end{figure*}

\subsection{Average shortest path length} \label{subsec:average-path-len-motivation}
The average shortest path length of a network measures the separation between nodes (on average) in term of the number of edges between nodes. Since this measure involves computing the separation between \textit{all nodes}, the network is required to be connected, i.e., there must exist a path from any node to any other node in the network. For a directed graph, the average shortest path length is the number of directed edges between nodes with the constraint that the distance should be measured along the direction of the edges. For undirected graphs, this is simply the average number of edges between nodes, irrespective of the directionality. This is often referred to as the all pairs shortest path (APSP), and is defined as,
\begin{align} \label{eq:apsp}
    l = \sum_{s,t \epsilon V}\frac{d(s,t)}{n(n-1)}
\end{align}
where $d(s,t)$ is the distance between nodes $s$ and $t$, and $n$ is the total number of nodes in the network.

To define connectivity for hypergraphs, we introduce two new concepts -- dual hypergraphs and linegraphs. First, the dual hypergraph $H^*$ of a hypergraph $H$ is a hypergraph with nodes and edges interchanged. Therefore, in an $H^*$, the nodes represent reactions and the hyperedges represent the set of molecules common between the nodes that it connects. Second, a linegraph $L(H)$ of a hypergraph $H$ is defined as a graph whose vertex set is the set of vertices of $H$ with two vertices adjacent and connected in $L(H)$ when their corresponding hyperedges have a non-empty intersection, i.e. they have common hyperedges (or reactions in our context). Therefore, a  hypergraph $H$ is said to be connected if its linegraph $L(H)$ is connected. A generalization of linegraphs is the s-linegraph where s (an integer, $\geq 1$) indicates the minimum size of the intersection, thus giving rise to s-linegraphs. Because of the duality property of hypergraphs, an equivalent linegraph $L(H^*)$ could be created for the dual hyeprgraph $H^*$ where the set of vertices represent hyperedges and adjacent vertices are connected if they have non-empty intersections, i.e. common molecules in our context. The s-linegraphs for the example set of reactions in Equation \ref{eq:example_reactions} for different values of $s$ is shown in Figure \ref{fig:slinegraphs} for $H$ and $H^*$. 

Now, for hypergraphs, the average shortest path length could be defined in the same manner as for dyadic graphs by computing the distance between nodes in an s-linegraph of H (known as s-distance). For our purpose, we generate the 1-linegraph and compute the average shortest 1-distance between the nodes using Equation \ref{eq:apsp}. 

Since the computation of the average shortest path length requires the graph to be connected, we find out the largest connected subcomponent both for the directed graph and the hypergraph and report their respective average shortest path lengths. For the example set of four reactions in Equation \ref{eq:example_reactions}, since both the directed graph representation and the hypergraphs's 1-linegraph representations are connected, their largest connected subcomponents are the same as their respective graphs (or hypergraphs). The average path lengths computed for the regular (undirected) graph and the hypergraph is show in Table \ref{tab:all_pairs_distance}.
\begin{table}[h]
\small
\centering
\caption{All pairs shortest distance for the example reactions}
\label{tab:all_pairs_distance}
\begin{tabular}{c|cc|cc}
\hline
\multirow{2}{*}{Node pairs} & \multicolumn{2}{c|}{Graph} & \multicolumn{2}{c}{Hypergraph} \\ \cline{2-5} 
                            & \multicolumn{2}{c|}{in}    & \multicolumn{2}{c}{s=1}           \\ \hline
$d_{A-B}$ & \multicolumn{2}{c|}{1} & \multicolumn{2}{c}{1}   \\
$d_{A-C}$ & \multicolumn{2}{c|}{1} & \multicolumn{2}{c}{1}   \\
$d_{A-D}$ & \multicolumn{2}{c|}{1} & \multicolumn{2}{c}{1}   \\
$d_{A-E}$ & \multicolumn{2}{c|}{2} & \multicolumn{2}{c}{1}   \\
$d_{B-C}$ & \multicolumn{2}{c|}{1} & \multicolumn{2}{c}{1}   \\
$d_{B-D}$ & \multicolumn{2}{c|}{2} & \multicolumn{2}{c}{2}   \\
$d_{B-E}$ & \multicolumn{2}{c|}{2} & \multicolumn{2}{c}{1}   \\
$d_{C-D}$ & \multicolumn{2}{c|}{1} & \multicolumn{2}{c}{1}   \\
$d_{C-E}$ & \multicolumn{2}{c|}{1} & \multicolumn{2}{c}{1}   \\
$d_{D-E}$ & \multicolumn{2}{c|}{1} & \multicolumn{2}{c}{1}   \\ \hline
\textbf{Average} & \multicolumn{2}{c|}{\textbf{1.3}} & \multicolumn{2}{c}{\textbf{1.1}} \\ \hline
\end{tabular}
\end{table}

\textit{Remark 3:} It is evident from the above table that in a hypergraph, the distances between nodes correspond exactly to the number of reactions that separate the nodes (or molecules), whereas in the case of a directed graph representation, the distance between nodes corresponds only to partial reactions separating the nodes and not the complete reactions.

\subsection{Assortativity} \label{subsec:assort-motivation}
\begin{figure}[h]
    \centering
    \subfigure[][]{{\includegraphics[width=0.46\linewidth]{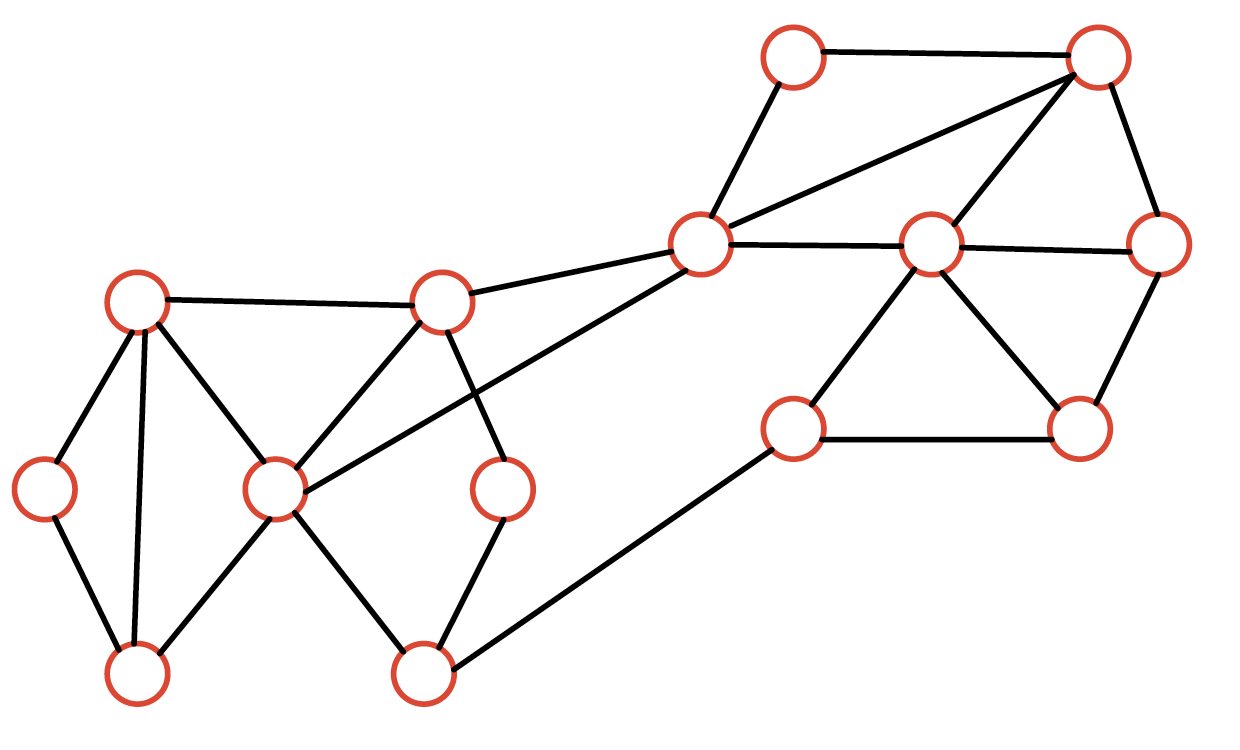}}}
    \quad
    \subfigure[][]{{\includegraphics[width=0.46\linewidth]{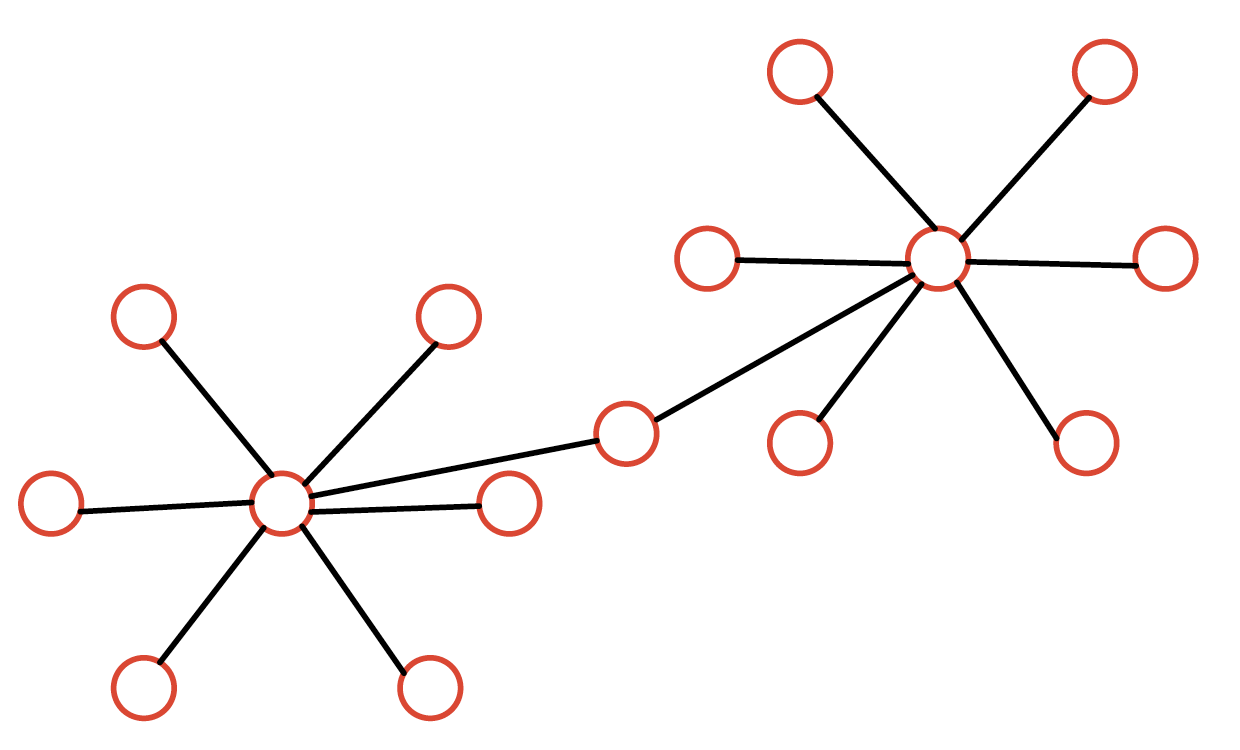}}}%
    \caption{Different mixing patterns (a) Assortative (b) Disassortative. Assortative networks have mixing patterns that arise due to nodes with similar degree connecting to other nodes with similar degrees, whereas disassortative networks are a result of mixing patterns where nodes with dissimilar degrees connect to each other}
    \label{fig:assortative_disassortative}%
\end{figure}

Assortativity is a measure of the \textit{mixing patterns} in networks that indicates the general mixing behavior of nodes with other nodes in the network to give rise to a bigger network. Assortativity is defined as the degree correlations between nodes, and therefore, the mixing pattern could either be assortative (positive correlation) or diassortative (negative correlation). The assortativity is often computed as the Pearson correlation coefficient between the degrees of a pair of nodes and takes values between -1 and 1 -- a network with an assortativity coefficient of -1 indicates a perfectly disassortative mixing, an assortativity coefficient of 1 points towards a perfectly assortative mixing, and an assortativity coefficient of 0 indicates a non assortative graph. Figure \ref{fig:assortative_disassortative} shows an example of assortative and disassortative networks. 

For a directed graph, the in-assortativity ($r_{in, in}$), out-assortativity ($r_{out, out}$), and in-out assortativity ($r_{out, in}$) measures the tendencies of nodes to connect with other nodes that have similar in-degrees, out-degrees, and out-in degrees, respectively. For $\alpha, \beta \in \{in,out\}$, the assortativity $r_{\alpha, \beta}$ for directed graphs is defined as
\begin{align} \label{eq:assortativity}
    r(\alpha, \beta) = \frac{\sum_i(j_i^\alpha - \bar{j^\alpha})(k_i^\beta - \bar{k^\beta})}{\sqrt{\sum_i(j_i^\alpha - \bar{j^\alpha})^2} \sqrt{\sum_i(k_i^\beta - \bar{k^\beta})^2}}
\end{align}
where $j_i^\alpha$ is the $\alpha$-degree of the source node for edge $i$,  $k_i^\beta$ is the $\beta$-degree of the target node for edge $i$, $\bar j^\alpha$ is the average $\alpha$-degree of source nodes, and  $\bar k^\alpha$ is the average $\beta$-degree of target nodes. For the annotated hypergraph, we define assortativity with respect to the roles (or annotations) in a manner similar to the directed graph representation in Equation \ref{eq:assortativity}, replacing the concept of edges with hyperedges and in-out degrees with role-specific (or annotation-specific) node degrees. 

For the example set of reactions in Equation \ref{eq:example_reactions}, the assortativity coefficients for the directed graph and the annotated hypergraph are reported in the Table \ref{tab:example_assort} below.

\begin{table}[h]
\small
\centering
\caption{Degree assortativity coefficients for the directed and hypergraph representations for the example set of four reactions}
\label{tab:example_assort}
\begin{tabular}{@{}ccc@{}}
\toprule
roles pair & directed & hypergraph \\ \midrule
p-p        & -0.19       & -0.43         \\
r-r        & -0.53    & -0.43      \\
r-p        & 0.27       & 0.15       \\ \bottomrule
\end{tabular}
\end{table}

\textit{Remark 4:} These assortativity values could be used to answer  questions such as -- how likely is it for products with high degree to connect to other products with high degrees, or how likely is it that the reactants would connect to other reactants of similar degrees (appear in reactions together), and so on.

\begin{table*}[t]
\centering
\caption{Network structure overview for the directed and hypergraph representation for the USPTO dataset}
\label{tab:uspto-network-props}
\resizebox{\textwidth}{!}{%
\begin{tabular}{ccc|cc|cc|cc}
 & \multicolumn{2}{c|}{all} & \multicolumn{2}{c|}{1976-1985} & \multicolumn{2}{c|}{1985-2005} & \multicolumn{2}{c}{after 2005} \\ \cline{2-9} 
      & \multicolumn{1}{c|}{graph} & hypergraph & graph & hypergraph & graph & hypergraph & graph & hypergraph \\ \hline
Num reactions & \multicolumn{2}{|c|}{487,724}                  &       \multicolumn{2}{c|}{69,692}           &       \multicolumn{2}{c|}{259,214}            &       \multicolumn{2}{c}{158,818}    \\  
Num (hyper)edges & \multicolumn{1}{|c}{1,245,533}      &  487,724 &    106,977     & 69,692      &   389,072        &  259,214      &    289,623          &     158,818  \\
Num nodes & \multicolumn{1}{|c}{440,207}      &  440,207 &    71,268     &  71,268     &     238,872       &    238,872   &     180,348       &   180,348                \\
\end{tabular}%
}
\end{table*}

\section{Network statistics on organic chemistry dataset} \label{sec:noc_results}
In this section, we study the network of organic chemistry through the lens of various network statistics defined in the previous section using a standard organic chemistry reactions database. The primary objective is to highlight the differences and similarities between the network statistics for the directed graph and the hypergraph representations. At the end of each section, we present chemistry insights that are drawn from such analyses along with the time-evolution of these properties.

\subsection{Dataset description} \label{subsec:data_descr}
The Jin's USPTO-reactions dataset \cite{jin2017predicting} derived from Lowe's text mining work \cite{lowe2014} for chemical reactions on the US patents office applications (1976-2016) is the primary dataset that we use to report and compare network statistics. We performed minimal preprocessing (removed incorrect, incomplete, and duplicate reactions) to allow for the network statistics to capture network properties without possibly losing information due to such preprocessing exercises. Along with information on reactants and products, the dataset also contained information on the year in which the reaction was reported, allowing us to investigate the time-evolution of the network properties. The final dataset contained 487,724 \textit{single-product reactions} containing information on participating reactants, major products of each reaction, and the year in which the reactions were reported.

Using this dataset, we construct directed graph and an annotated hypergraph-based networks of organic chemistry. The directed graph representation was constructed using the all-to-all node connectivity for each reaction. The other wiring possibilities are one-to-one or many-to-one but it has been shown previously that the actual connectivity pattern does not change the network structure and properties \cite{fialkowski2005architecture, jacob2018statistics}. The annotated hypergraph, on the other hand, represents all the reactants and products as part of the same hyperedge with node annotations based on
\begin{itemize}
    \item reaction roles: `reactant' or `product'
    \item relative length of SMILES strings \textit{in a reaction} with respect to the median SMILES length per reaction: `SMILES\_short', `SMILES\_medium', `SMILES\_long'
    \item molecular weight \textit{across the entire dataset}: `molwt\_light', `molwt\_medium', `molwt\_heavy'
\end{itemize}

To perform an analysis of the time-evolution of network properties over different stages of chemistry research, we split the data into three time regimes -- regime 1 with reactions reported from 1976 to 1985, regime 2 with reactions reported after 1985 until 2005, and regime 3 with reactions reported from 2005 until 2016. An overview of the directed graph and hypergraph representation obtained using the entire dataset and also using dataset in the three time-regimes is presented in Table \ref{tab:uspto-network-props}.

\textit{Remark 5:} Note that in the case of the hypergraph, the number of hyperedges exactly equals the number of reactions in the dataset, whereas for the graph representation, the number of edges is much higher. Of course, the number of nodes remain the same in both the representations since each node corresponds to a unique molecule in both the representations.


\subsection{Degree distributions} \label{subsec:degree-distr-results}
\subsubsection{Degree distribution comparison}
\begin{figure}[h]
    \centering
    \includegraphics[width=0.98\linewidth]{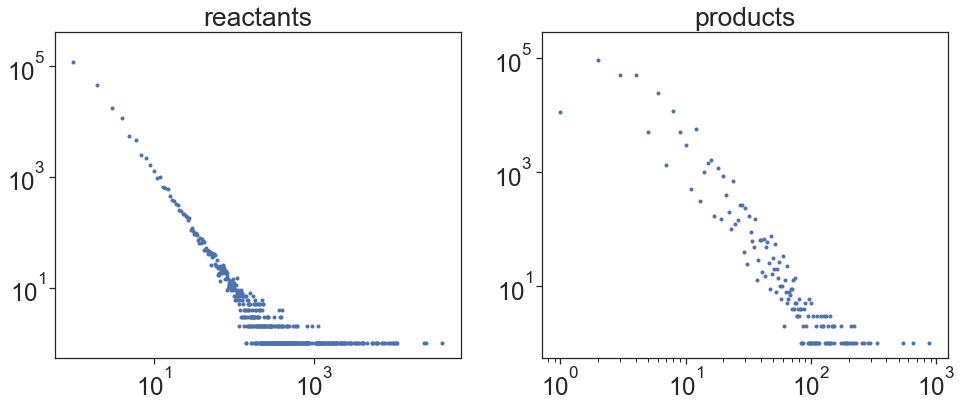}
    \caption{Degree distributions for outgoing (reactants) and incoming (product) edges in a directed graph }
    \label{fig:degree-distrs-directed}
\end{figure}

\begin{figure}[h]
    \centering
    \includegraphics[width=0.49\textwidth]{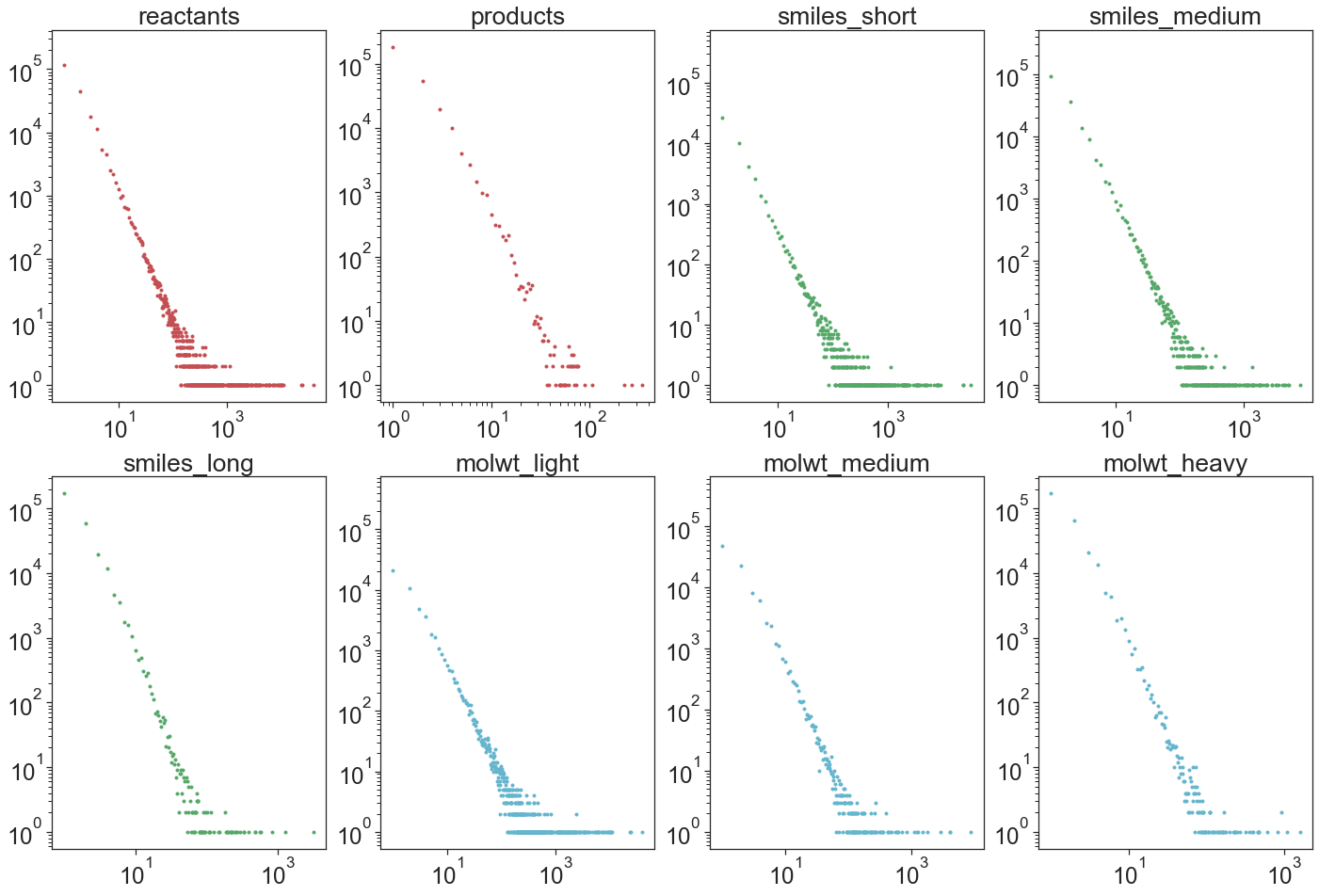}
    \caption{Degree distributions for the various hypergraph node-annotations (or roles)}
    \label{fig:additional-annotations-degree-distrs}
\end{figure}

We first compare the degree distributions of both the incoming and outgoing degrees for the directed graph and annotated hypergraph representations. Recall from Section \ref{subsec:degree-distr-motivation} that for the annotated hypergraph, the incoming degree is the same as the node-degree for annotation `product' and the outgoing degree is the same as the node-degree for annotation `reactant'. The degree distributions for the directed graph and for the various annotations in the hypergraph (based on reaction roles, relative SMILES length, and molecular weights as defined in the foregoing section) are presented respectively in Figures \ref{fig:degree-distrs-directed} and \ref{fig:additional-annotations-degree-distrs}. 


\textit{Remark 6:} Note that since our dataset only contains single-product reactions, the outgoing degree distribution (reactants) is the same in both representations, and only the incoming (product) degree distributions differ. This is because the directed graph representation for a reaction would have as many incoming edges for the product as the number of reactants whereas in the hypergraph representation the product would have just one incoming edge.

\subsubsection{Power law fit for degree distributions}
\begin{figure}[h]
    \centering
    \subfigure[][]{{\includegraphics[width=0.47\linewidth]{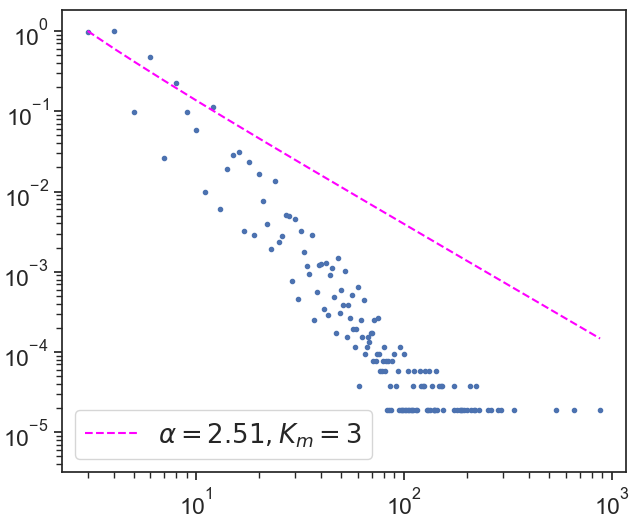}}}
    \quad
    \subfigure[][]{{\includegraphics[width=0.47\linewidth]{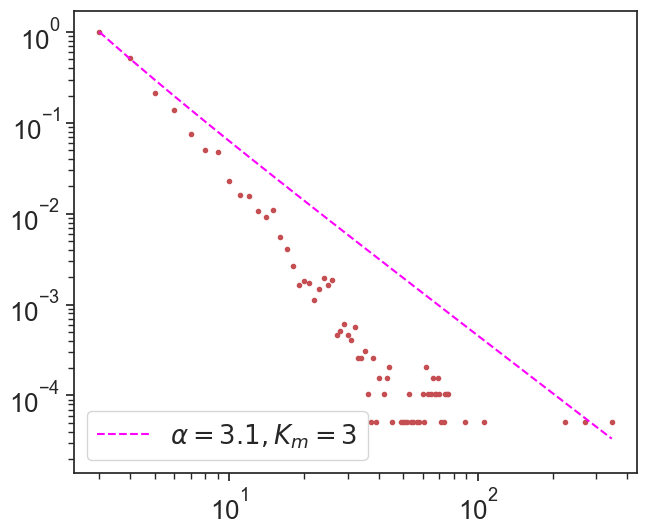}}}%
    \caption{Scale-free distribution fit on incoming (products) degrees for (a) Directed graph (b) Hypergraph; $K_m$ is the minimum degree cutoff threshold that is required as a hyperparameter in the powerlaw package}
    \label{fig:powerlaw-incoming-dir-hyper}%
\end{figure}

A visual inspection of the degree distributions indicates a possible power law distribution, which is defined as
\begin{align}\label{eq:powerlaw}
    p(k) \propto k^{-\alpha}
\end{align}
where $p(.)$ is the degree distribution, $k$ is the degree, and $\alpha$ is the scale-free or power law distribution parameter. The existence of a power law distribution points towards an underlying network structure known as the scale-free network structure \cite{barabasi2003scale}, ubiquitous in real-world networks that often results in `small-world' behavior. We perform a mathematically rigorous fit to ensure the existence of a power law using the powerlaw package in Python and estimate the underlying scale-free distribution parameter. The power law fit for the incoming degrees (products) for the directed graph and the hypergraph-based network of organic chemistry are shown in Figure \ref{fig:powerlaw-incoming-dir-hyper}. 

We observe that the degree distributions for both the directed graph and hypergraph incoming degrees could be assumed to be coming from a power law distribution, thus pointing towards an underlying scale-free network structure, agreeing with several other studies that have shown that chemistry networks exhibit a scale-free or small-world behavior \cite{jacob2018statistics, fialkowski2005architecture, grzybowski2009wired}. However, the scale-free parameter, $\alpha$ differs in both the cases -- $\alpha$ is 2.51 for the directed graph (close to 2.7 reported in \cite{grzybowski2009wired, fialkowski2005architecture} on another reactions dataset) and 3.1 for the hypergraph.

In order to ascertain the difference in $\alpha$ values for the degree distributions, we estimate the scale-free parameter by randomly sub-sampling different fractions of the network in a step-forward manner in time, i.e., by utilizing the reaction year information, we sample reactions starting from 1976 sequentially sampling additional reactions from the following years. We sub-sample $0.1 - 1.0$ fraction of the network in steps of $0.1$ and repeat this 10 times to perform bootstrapping and compute the deviations in $\alpha$. The results are presented in Table \ref{tab:alpha_values_fracs}. It is clear from the table that the scale-free distribution is indeed different in the two representations and remains the same irrespective of the fraction of network sub-sampled for estimating the distribution.

\begin{table}[h]
\centering
\caption{Scale-free distribution parameter values, $\alpha$, for different fractions of the network sampled using step-forward sampling in time using 10 bootstrapped samples for each fraction}
\label{tab:alpha_values_fracs}
\resizebox{0.35\textwidth}{!}{%
\begin{tabular}{@{}c|cc|cc@{}}
\toprule
frac & \multicolumn{2}{c|}{graph $\alpha$} & \multicolumn{2}{c}{hyeprgraph $\alpha$} \\ \cmidrule(l){2-5} 
     & mean          & std              & mean             & std                \\ \midrule
0.1  & 2.54          & 0.0009           & 3.1              & 0.0014             \\
0.2  & 2.54          & 0.0003           & 3.18             & 0.0013             \\
0.3  & 2.48          & 0.0008           & 3.13             & 0.0021             \\
0.4  & 2.48          & 0.0005           & 3.1              & 0.0013             \\
0.5  & 2.47          & 0.0001           & 3.02             & 0.0003             \\
0.6  & 2.48          & 0.0004           & 3.03             & 0.0014             \\
0.7  & 2.49          & 0.0005           & 3.04             & 0.0014             \\
0.8  & 2.5           & 0.0003           & 3.07             & 0.0005             \\
0.9  & 2.51          & 0.0001           & 3.1              & 0.0001             \\
1.0  & 2.52          & 0.00014          & 2.97             & 0.0034             \\ \bottomrule
\end{tabular}%
}
\end{table}

\subsubsection{Time-evolution of scale-free network property}
\begin{figure}[h]
    \centering
    \subfigure[][Directed graph incoming degrees]{{\includegraphics[width=0.65\linewidth]{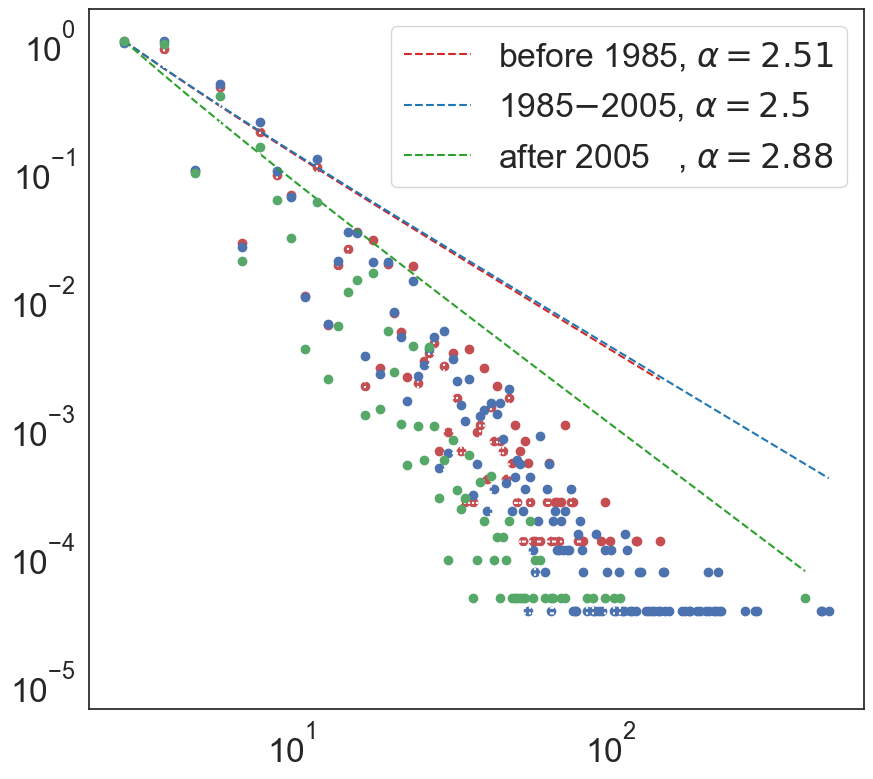}}}\\
    \subfigure[][Hypergraph `product' degrees]{{\includegraphics[width=0.65\linewidth]{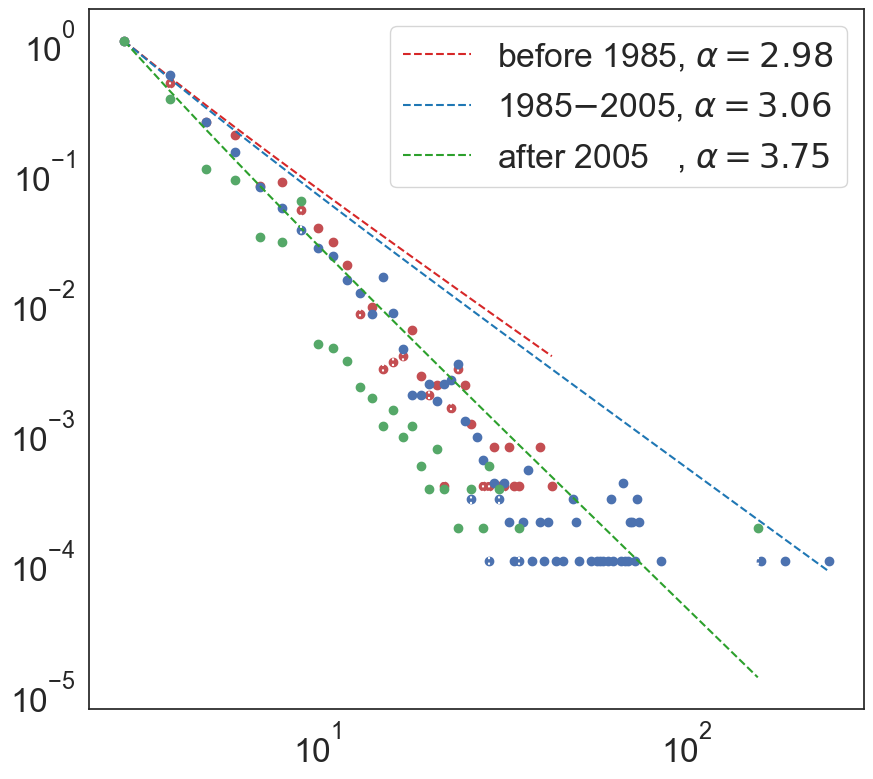}}}
    \caption{Scale-free fit for reactions reported in the three regimes with estimated $\alpha$ in inset}
    \label{fig:deg-distr-time-evolution}%
\end{figure}

Next, we study the time-evolution of the scale-free parameter $\alpha$ by computing it across the three time regimes -- before 1985, 1985 -- 2005, and after 2005. The degree distributions, power law fit, and the estimated $\alpha$ values for the power law fit are show in Figure \ref{fig:deg-distr-time-evolution}.
We observe that the scale-free parameter $\alpha$ has been increasing over the years with significant increase post 2005, pointing towards accelerated growth nature of the hypergraph network \cite{albert2002statistical} and a similar observation has been made on reactions dataset in \cite{grzybowski2009wired}. The accelerated growth of the network of chemistry is also evident from the average path length analysis presented in Section \ref{subsec:average-path-len-results}. 

\subsubsection{Inferences from degree distributions analysis}
First, we observe that the degree distributions in both the cases follow a scale-free distribution, pointing towards an underlying mechanism of `preferential attachment' or `preferential linking' where new nodes attach to existing nodes in the network with probability proportional to their connectivity or node degrees. Mathematically, preferential attachment is characterized by
\begin{equation}
    \Pi(k) \sim k^c
\end{equation}
where $\Pi(k)$ is the probability of a new node attaching to an existing node with degree $k$, and $c$ is a constant controlling the degree of non-linearity in preferential attachment. This expression translates to the inference that chemistry growth is largely driven by a relatively small set of highly important molecules that are highly connected (higher degree, $k$) and they have a higher likelihood of playing a central role in the discovery of new molecules or reactions because of the underlying phenomenon of preferential linking.

Second, for the directed and hypergraph representations, the parameter characterizing the scale-free distributions is higher for the latter. This could be related to chemistry by looking at the concept of `initial attractiveness' in scale-free networks that assigns a non-zero probability of connecting to an isolated node, given by
\begin{equation}
    \Pi(k) = A + k^c
\end{equation}
which ensures that, for non-zero values of $A$,  $\Pi(k)\neq0$ for disconnected nodes. The presence of A in the expression for preferential attachment $\Pi(k)$ does not affect the scale-free structure of the network but has a direct-impact on the $\alpha$ parameter as,
\begin{equation} \label{eq:initial-attract}
    \alpha = 2 + (k1+A)/k2
\end{equation}
where $k1,k2$ are constants with values depending on the underlying generating model and A characterizes the initial attractiveness of nodes. Thus, it could be inferred from Equation \ref{eq:initial-attract} that the initial attractiveness based on the hypergraph representation is higher than that of the directed-graph representation since the former has a higher $\alpha$ of 3.1 characterizing the scale-free distribution compared with $\alpha$ of 2.51 for directed-graph representation. Moreover, the gradually increasing $\alpha$ values for the scale-free distribution in both directed and hypergraph representations indicates that the initial attractiveness has been increasing over time, with the trend being much more evident in the latter representation where $\alpha$ grew from 2.98 in regime 1 to 3.75 in regime 3. 

Third, a higher initial attractiveness translates to a higher likelihood of discovering new connections (or reactions) to isolated nodes (rare or complex molecules). Since the initial attractiveness is the highest and much different in regime 3 (after 2005) than the other two regimes, it could be inferred that in the recent years, there has been an emphasis on the rewiring of existing reactions to create connections between previously disconnected nodes, or the synthesis of rarer molecules. It will become clear from the analysis in the next section on average shortest path length that the major driver of chemistry evolution in the recent years is the rewiring of existing reactions.

\subsection{Average path length} \label{subsec:average-path-len-results}
\subsubsection{Average path length comparison}
The average separation between the molecules (vertices) in terms of number of reactions (edges) is captured by the average path length of the network. We compute the average path length on the largest connected subgraph for both the representations. Recall from the Section \ref{subsec:average-path-len-motivation} that for the hypergraph, in order to make a one-to-one comparison, we choose $s=1$ to generate a 1-linegraph and compute the 1-distance between nodes to compute the average shortest path length for the hypergraph. The average shortest path lengths for the largest connected subgraph obtained for the two representations for different fraction of nodes sampled from the entire dataset using step-forward sampling is shown in Table \ref{tab:entire-graph-apsp}.
\begin{table}[H]
\centering
\caption{All pairs shortest path (or APSP) on the entire dataset}
\label{tab:entire-graph-apsp}
\begin{tabular}{@{}c|c|cc|cc@{}}
\toprule
\multirow{2}{*}{Fraction} & \multirow{2}{*}{Reactions} & \multicolumn{2}{c|}{Directed} & \multicolumn{2}{c}{Hypergraph} \\ \cmidrule(l){3-6} 
      &         & nodes   & APSP   & nodes  & APSP    \\ \midrule
1\%   & 4,877   & 7,528   & 6.62 & 7,516  & 3.99 \\
5\%   & 24,386  & 26,222  & 5.98 & 26,176 & 3.75 \\
10\%  & 48,772  & 47,043  & 5.69 & 47,069 & 3.64 \\
20\%  & 97,544  & 91,903  & 5.36 & 91,870 & 3.52 \\
50\%  & 243,862 & 209,790 & 5.16 &    \textcolor{gray}{209,790}    &  \textcolor{gray}{3.37}\footnotemark[2]   \\
\textbf{100\%} & \textbf{487,724} & \textbf{411,396} & \textbf{5.11} &   \textbf{\textcolor{gray}{411,396}}   &  \textbf{\textcolor{gray}{3.25}\footnotemark[2]}    \\ \bottomrule
\end{tabular}%
\end{table}

\footnotetext[2]{\label{f1}extrapolated values since the network size was prohibitively large for the hypernetX package in python with no C-optimized libraries}
\subsubsection{Time evolution of Average path length}
\begin{figure}[h]
    \centering
    \subfigure[][Graph]{{\includegraphics[width=0.65\linewidth]{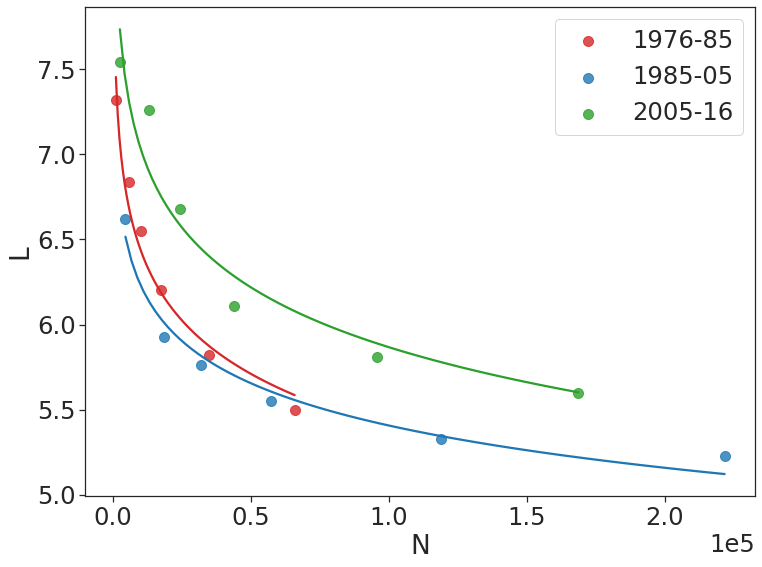}}}\\
    \subfigure[][Hypergraph]{{\includegraphics[width=0.65\linewidth]{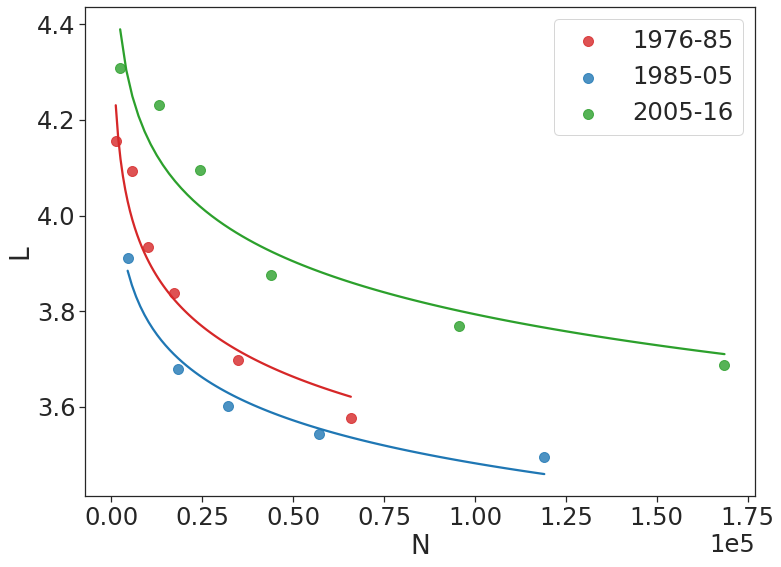}}}
    \caption{Average shortest path lengths for various regimes as function of the number of nodes in the sub-sampled graph}
    \label{fig:pathlen_avg}%
\end{figure}

Similar to the degree distribution analysis, we study the time-evolution of the average path lengths of the networks in the three time regimes. The average path length as a function of the number of nodes in the network using time-based step-forward sampling is shown in Figure \ref{fig:pathlen_avg} for both the representations for different fractions of the networks, namely $1\%, 5\%, 10\%, 20\%, 50\%$ and $100\%$ of the network in each regime. 
We observe that across both the representations, the average path length between the nodes decreases exponentially as the number of nodes in the networks is increased. Moreover, in both the cases, the average path lengths for the time regimes 1 and 2 are very similar to each other but the average path length for regime 3 is significantly higher than those in other two across all values of N. The phenomenon of decreasing path length as number of nodes is increased has been reported in the literature as network densification \cite{leskovec2007graph} where the network grows more and more dense over time -- this makes sense for the USPTO dataset containing patented reactions where the nodes are mostly sparsely connected and they get more connected over time after either new reactions discovered, or existing nodes become more connected.

In terms of differences, the average path length is much smaller for the hypergraph representation than for the directed graph. This could be an outcome of the frugal representation of hypergraphs where number of edges is the same as number of reactions but that is not the case with graphs. This is one of the major advantages of using hypergraphs -- the edges-based analysis has a one to one correspondence with reactions-based analysis, meaning that the separation in terms of hyperedges between nodes corresponds exactly to the separation between molecules in terms of reactions. Therefore, the average separation between nodes in a hypergraph not only differs from a dyadic graph representation, but the separation corresponds to the number of reactions (on average) the separate the nodes in the network.

\subsubsection{Inferences from average path lengths analysis}
First, we observe that as expected, the average shortest path length for the directed graph representation is higher than that of the hypergraph representation. This again is an artifact of the directed graph representation which introduces several additional edges for each reaction depending on the number or reactants and products per reaction. On the other hand, in the hypergraph representation, since each hyperedge connects all the molecules taking part in a reaction using a single hyperedge, the separation exactly equals the number of reactions separating any two given nodes. Thus, in the directed graph average shortest path length for the entire network is 5.11 while in the hypergraph it is 3.25.

Second, the average all pairs shortest distance for the hypergraph could be interpreted as separation between nodes (or molecules) in terms of number of reactions. Recall that since each hyperedge corresponds to a unique reaction, there exists a one to one mapping between the number of reactions and the number of hyperedges separating the molecules. Thus, the hypergraph network of organic chemistry indicates that the network of organic chemistry is much more compact than previously understood with nearly 3.25 degrees of separation between molecules, pointing towards an even stronger small-world nature than previously observed with five degrees of separation \cite{jacob2018statistics}. 

Third, in both the cases, the time-evolution of the network suggests that over time, network densification takes place primarily due to the creation of links between existing nodes in the network rather than by the addition (or discovery) of new nodes. A characteristic of network densification is shrinking diameter \cite{leskovec2007graph}, i.e., the average separation between nodes decreases as the network grows, similar to the exponential decrease in average shortest path length reported in Table \ref{tab:entire-graph-apsp} and Figure \ref{fig:pathlen_avg}. This phenomenon is observed for both the representations and across time-regimes, pointing towards an underlying process causing the densification. There exist models for explaining such densification such as the community guided attachment similar to preferential attachment but at a bigger community (or cluster) level with separation between the communities \cite{leskovec2007graph}. However, the exact quantitative model guiding densification in reaction chemistry network needs further studies. Nevertheless, densification suggests that chemistry has been evolving mostly based on the rewiring of existing reactions (edges) rather than the discovery of completely new molecules (nodes addition), that has brought the molecules closer to each other over time. This is intuitive for the reaction patents dataset that we worked with since most molecules are initially well separated given that they are patented molecules/reactions which get more connected (reachable) over time due to the discovery of new reactions over the years. 

Third, the time-evolution analysis of the average shortest path length in Figure \ref{fig:pathlen_avg} suggests that in regime 1 and 2, the average separation between molecules was nearly the same for a given number of nodes, $N$ in the network. However, in regime 3, there was a significant upward shift of average separation across all values of $N$. This suggests that the time-regime post 2005 is characterized by the discovery of complex chemistry leading to the synthesis of molecules via complex routes that has led to the increase in their average separation, possibly due to significant advances in computational capabilities around this time. This increase in average separation is more evident in the hypergraph representation than the directed-graph representation.


\subsection{Assortativity} \label{subsec:assort-results}
\subsubsection{Assortativity comparison}
To understand the mixing patterns of nodes in the two network representations, we compute the assortativity values between different node-type (or role) combinations -- `in' and `out' degree roles for directed graphs and pairwise roles-based node degrees for the annotated hypergraph. Table \ref{tab:assort_entire} shows the assortativity values for the two representations on the entire network. The assortativity values for the two representations agree qualitatively with each other but differ in terms of their relative strengths. From Table \ref{tab:assort_entire}, we see that in the hypergraph representation, the reactant nodes exhibit strong assortative mixing (out-out). On the other hand, the product-product and reactant-product exhibit very weakly assortative or non-assortative behavior pointing towards a lack of degree correlation between such nodes.
\begin{table}[h]
\caption{Assortativity values on the entire dataset}
\label{tab:assort_entire}
\centering
\begin{tabular}{@{}ccc@{}}
\toprule
\multicolumn{1}{l}{{node-pairs}} & \multicolumn{1}{l}{{directed graph}} & \multicolumn{1}{l}{{hypergraph}} \\ \midrule
in-in   & 0.0107 &  0.0734 \\
out-out  & 0.0049 &  0.1159  \\
out-in  & 0.0187 &  0.0032 \\ \bottomrule
\end{tabular}%
\end{table}
Owing to the flexibility offered by annotations in the hypergraph, we computed additional assortativity between node roles of reactant/product with roles based on molecular weights and relative SMILES length of molecules, as shown in Tables \ref{tab:assort_entire_molwt} and \ref{tab:assort_entire_smiles_len}. We observe that reactant-MW${}_{\text{light}}$ and reactant-SMILES${}_{\text{short}}$ exhibit strong assortative mixing whereas this is not usually the case with other node-role pairs.

\begin{table}[H]
\caption{Hypergraph assortativity between reactant \& product roles with roles based on molecular weights}
\label{tab:assort_entire_molwt}
\centering \small
\begin{tabular}{@{}llll@{}}
\toprule
         & MW${}_{light}$ & MW${}_{medium}$ & MW${}_{heavy}$ \\ \midrule
reactant &     0.1337         &      0.0074         &    0.0061          \\
product  &     0.0119         &     0.0003          &   0.0004           \\ \bottomrule
\end{tabular}%
\end{table}
\begin{table}[H]
\caption{Hypergraph assortativity between reactant \& product roles with roles based on relative SMILES lengths}
\label{tab:assort_entire_smiles_len}
\centering \small
\begin{tabular}{@{}llll@{}}
\toprule
         & SMILES${}_{short}$& SMILES${}_{medium}$ & SMILES${}_{long}$ \\ \midrule
reactant &      0.1782        &       0.0713        &   0.0015           \\
product  &        0.0237      &       0.0083        &     -0.0009         \\ \bottomrule
\end{tabular}%
\end{table}

\subsubsection{Time evolution of assortativity}
To study the evolution of mixing patterns in the network over time, we study the time-evolution of assortativity during the three time regimes. The assortativity values for the in-in, out-out, and out-in node-role pairs are shown in Table \ref{tab:assort_timebased} below.
\begin{table*}[h]
\caption{Time evolution of assortativity for the directed graph and hypergraph for various node-role pairs}
\label{tab:assort_timebased}
\centering
\begin{tabular}{@{}cccc|ccc@{}}
\toprule
\multirow{2}{*}{node-pairs} & \multicolumn{3}{c|}{Directed graph}  & \multicolumn{3}{c}{Hypergraph}       \\ \cmidrule(l){2-7} 
                            & before 1985 & 1985-2005 & after 2005 & before 1985 & 1985-2005 & after 2005 \\ \midrule
in-in   & 0.0630 & 0.0157 & 0.0175 & 0.3623 & 0.2302 & 0.1674 \\
out-out & 0.0047 & 0.0042 & 0.0069 & 0.2368 & 0.2241 & 0.2259 \\
out-in  & 0.0274 & 0.0257 & 0.0286 & 0.0005 & 0.0057 & 0.0076 \\ \bottomrule
\end{tabular}%
\end{table*}

\begin{figure}[h]
    \centering
    \subfigure[][\scriptsize{Assortativity between MW and reactant/product roles. Reactants show assortative mixing with nodes with MW${}_{light}$ across time regimes, whereas products show assortative mixing with MW${}_{heavy}$ before 1985.}]{{\includegraphics[width=0.98\linewidth]{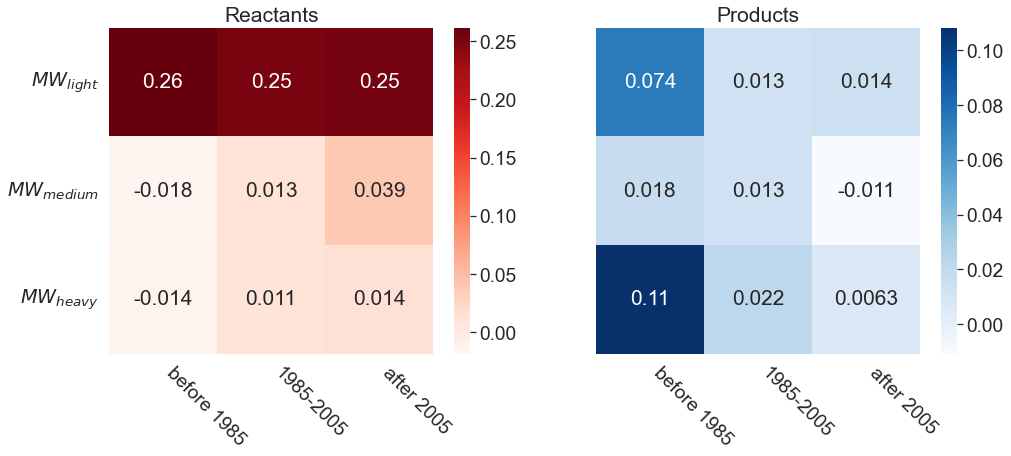}}}\\
    \hspace{-1em}
    \subfigure[][\scriptsize{Assortativity between relative SMILES length and reactant/product roles. Reactants show assortative mixing with nodes with SMILES${}_{short}$ and SMILES${}_{medium}$ across time regimes, whereas products show assortative mixing with SMILES${}_{short}$ and SMILES${}_{long}$ before 1985 and with SMILES${}_{medium}$ from 1985-2005.}]{{\includegraphics[width=0.98\linewidth]{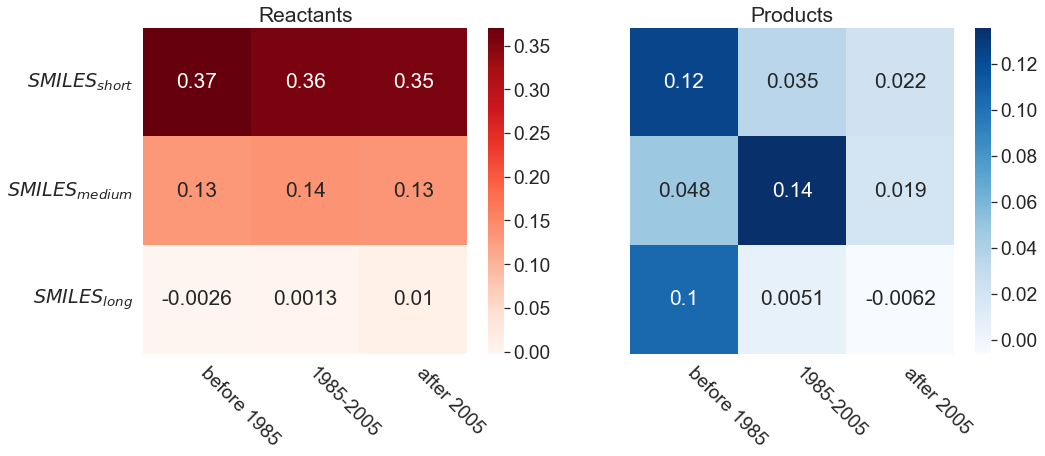}}}\\
    \caption{Time evolution of assortativity for reactants and products with respect to additional node annotations (or roles)}
    \label{fig:assort-addnl-time-evolution}%
\end{figure}

We observe from the table above that the directed-graph representation does not show any strong trend in various assortativity values, an observation also reported in \cite{jacob2018statistics}. On the other hand, the hypergraph representation shows a decreasing assortativity of in-in nodes over time and an increasing assortativity of out-in nodes. A further analysis on additional assortativity values with different node-role pairs reveal additional trends as shown in Figure \ref{fig:assort-addnl-time-evolution}. We observed that reactants show assortative mixing with nodes with MW${}_{light}$ across time regimes, whereas products show assortative mixing with MW${}_{heavy}$ before 1985. Similarly, we also observe that reactants show assortative mixing with nodes with SMILES${}_{short}$ and SMILES${}_{medium}$ across time regimes, whereas products show assortative mixing with SMILES${}_{short}$ and SMILES${}_{long}$ before 1985 and with SMILES${}_{medium}$ from 1985-2005.

\subsubsection{Inferences from assortativity analysis}
The assortativity analysis highlights another limitation of the directed-graph representation in terms of obscuring the underlying network characteristics induced by the network wiring scheme. Based on the observations in Table \ref{tab:assort_entire} for the directed-graph representation, it appears that the network is non assortative or very weakly assortative with respect to all the node-role pairs. It was shown in \cite{jacob2018statistics} that this is an artifact of the network preprocessing and the assortativity values change drastically if one chooses to perform network preprocessing to remove parallel edges. On the contrary, the hypergraph representation shows that the network is assortative with respect to certain node-role pairs such as out-out degree assortativity indicating that commonly used reactants tend to take part in reactions together. 

Second, due to the flexibility of the hypergraph representation in terms of allowing additional node annotations, we performed additional assortativity analysis with respect to different node-role pairs as shown in Tables \ref{tab:assort_entire_molwt} and \ref{tab:assort_entire_smiles_len}. It was observed that reactants are assortative with molecules of light molecular weight and relatively short/medium SMILES length, highlighting the mixing patterns of reactant nodes in the network. Products, on the other hand, seem to be non-assortative with these properties, thus highlighting the wide spectrum of products with varying degrees of complexity present in the dataset. 

An analysis of the time-evolution of assortativity presented in Table \ref{tab:assort_timebased} shows node-mixing trend across time-regimes, with no clear trend in assortativity for directed-graphs. However, from the hypergraph representation it is observed that the reactants exhibit assortative mixing at nearly the same level across time regimes, whereas the products show a decreasing assortativity over time. The latter points towards the general trend in earlier years (regime 1) to discover several different routes for synthesizing a given molecule, which has been decreasing over the years (but still significant) due to the  synthesis of new products molecules with different chemistry.

Finally, based on the time-evolution of assortativity with respect to additional node annotations in Figure \ref{fig:assort-addnl-time-evolution}, we observe that reactants are assortative at the same level with heavy molecular weight as well as relative molecular complexity across time regimes, with decreasing assortativity as the molecular weight or complexity is increased. Products on the other hand, show a positive assortativity before 1985 with heavy and complex molecules, in regime 2 assortative with medium complexity, and non-assortative in regime 3 with all roles. The latter indicates towards the diversity of products synthesized in the recent years.

\section{Additional hypergraph statistics} \label{sec:additional-stats-proj}
Even though many dyadic network properties could also be defined equivalently for hypergraphs, sometimes it is necessary to work with the directed graph framework for reasons among -- interpretability from a traditional graph-theoretic standpoint, easy availability of tools for computation of dyadic properties, or aversion towards adopting hypergraphs due to their seemingly high complexity. The annotated hypergraph could, therefore, be projected as a directed graph with edge-weights defined using a role-interaction kernel \cite{chodrow2020annotated}. The role-interaction kernel defines the mapping of the annotated hypergraph to a projected-directed graph, that maps various nodes to annotations in the hypergraph using weighted edges. We work with the following three kernels:
\begin{itemize}
    \item $R1 = \begin{bmatrix} 1 & 0 \\ 0 & 0 \end{bmatrix}$: each hyperedge split into multiple weighted directed edges from reactants to products each with weight 1; emphasis is on forward reactions only
    \item $R2 = \begin{bmatrix} 0 & 0.75 \\ 0.25 & 0 \end{bmatrix}$: each hyperedge split into multiple weighted directed edges with directed edges from reactants to products with weight 0.75 and also directed edges in the reverse direction (from products to reactants) with weight 0.25; unequal emphasis on forward and inverse reactions
    \item $R3 = \begin{bmatrix} 0 & 0 \\ 0 & 1 \end{bmatrix}$: each hyperedge split into multiple weighted directed edges in the reverse direction (from products to reactants) each with weight 1; emphasis on inverse reactions only
\end{itemize}
Using such projected dyadic graphs, we perform two additional studies on the entire network -- first, a PageRank \cite{page1999pagerank} analysis of reaction nodes to identify the most important molecules, and second, a graph-based community-detection (or clustering) \cite{traag2019louvain} to identify clusters in the reaction networks based on their connectivity patterns.

\subsection{PageRank analysis} \label{subsec:pagerank-results}
The PageRank algorithm was originally proposed for ranking of webpages on the internet \cite{page1999pagerank} based on the number and quality of links to webpages and is based on a random-surfer model that performs random walks along incoming and outgoing edges from webpages. A page that has a higher likelihood of being visited by a random surfer is therefore considered more important by PageRank, thus  requiring both higher connectivity as well as connectivity to other important webpages for higher PageRank. 

Extending the idea of PageRank to chemical reactions and molecules, we could find the set of molecules that are most important based on their connectivity (high reactivity) as well as their connectivity to other important molecules (chemical importance due to ease of synthesizability or criticality for other compounds). Thus, a molecule with high PageRank in a network of chemical reactions should be crucial both from a reactivity/synthesizability as well as reachability/criticality standpoint. In contrast, a molecule with merely the highest degree does not say much about the molecule except that the molecule participates in many reactions. 

\begin{figure*}[h]
    \centering
    \includegraphics[width=.75\textwidth]{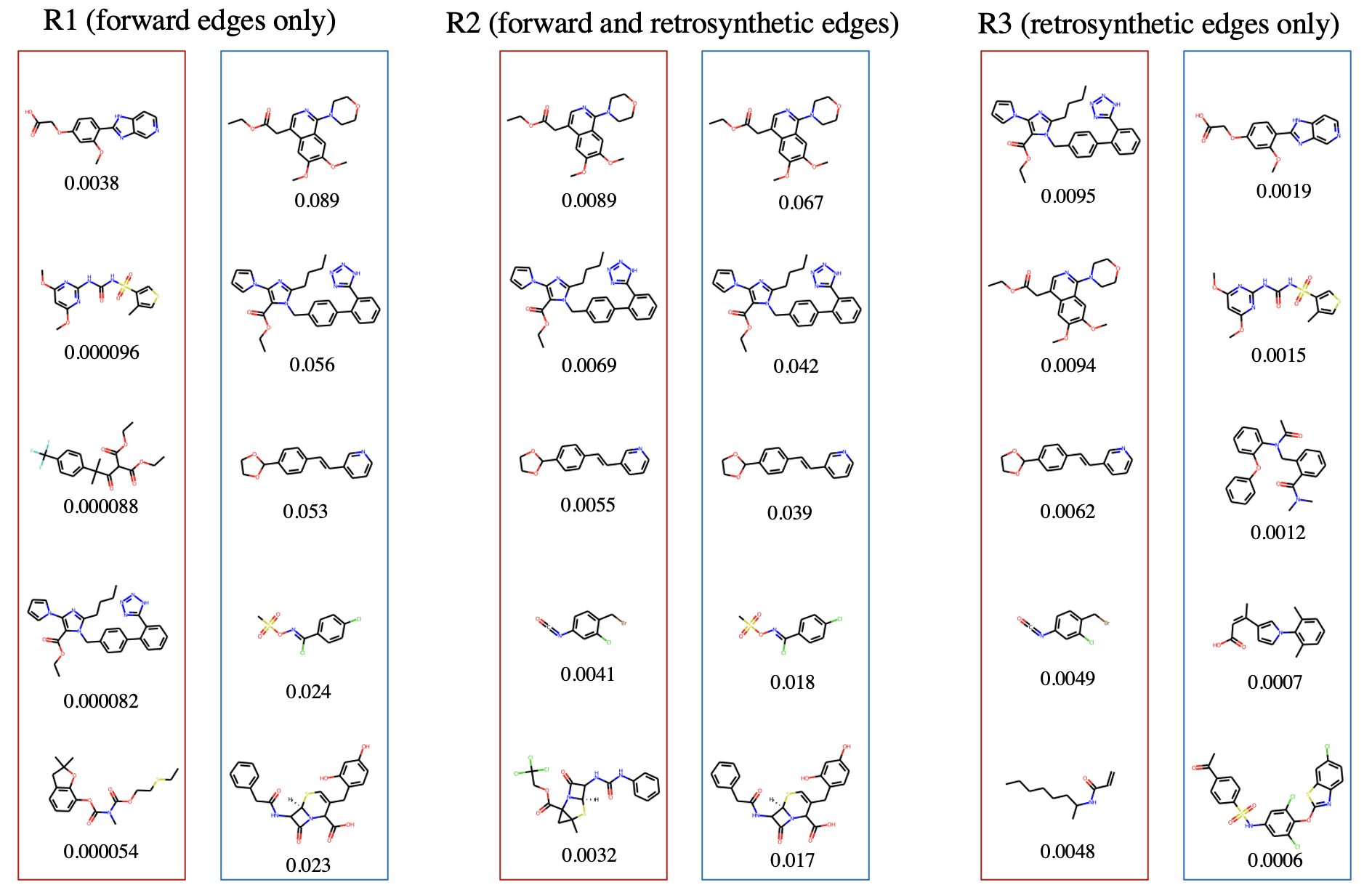}
    \caption{PageRank and degree centrality analysis for the three role interaction kernels with $R1 = \bigg[\begin{matrix}0 & 1\\ 0 & 0\end{matrix}\bigg]$, $R2 = \bigg[\begin{matrix}0 & 0.75\\ 0.25 & 0\end{matrix}\bigg]$, and $R3 = \bigg[\begin{matrix}0 & 0\\ 1 & 0\end{matrix}\bigg]$  corresponding to forward edges only, forward and retrosynthetics edges, and retrosynthetic edges only.} 
    \label{fig:deg_pr_analysis}
\end{figure*}

Using the three role-interaction kernels -- R1, R2 and R3 defined above, we compute the PageRank and degree centrality of nodes in the resulting network defined as $d_v/d_{max}$ where $d_v$ is the degree of node $v$ and $d_{max}$ is the maximum degree across all nodes in the network. Since PageRank and degree centrality are two different measures, their absolute values should not be compared and only the relative values or ranked order of molecules should be compared. The top-5 molecules based on PageRank and degree centralities computed using the weighted directed reaction networks obtained using different role-interaction kernels are shown in Figure \ref{fig:deg_pr_analysis}. 

Based on the above ranked order of molecules, we first observe that the molecules that are important from a PageRank standpoint are not the same as those important from a degree centrality standpoint. Second, across the role interaction kernels, the ranked order changes, i.e., molecules critical based on R1 kernel-based projection (forward edges) of hypergraph differ from those based on R3 kernel-based projection (retrosynthetic edges). This highlights the flexibility of the hypergraph reaction representation in incorporating custom importance for forward and retrosynthetic reaction directions through role-interaction kernels. Such an analysis of molecular importance in a reaction network would have application in optimizing reaction networks, designing robust supply chain networks, and performing efficient product design.

\subsection{Community detection analysis} \label{subsec:community-detect-results}
To study the formation of communities or clusters in the reaction network based on the mutual connectivity patterns and node-densities, we perform graph-based clustering on the network of reactions. We use the Leiden algorithm \cite{traag2019louvain} to perform optimal graph partitioning that results in well-connected set of dense nodes in the network (called communities) and is a suitable algorithm for weighted, directed networks. For this study, we use the R2 role-interaction kernel to preserve both forward and retrosynthetic edges but with unequal weights in the network. The applications of such a graph-based community detection exercise is to get a general sense of the distribution and connectivity patterns of reactions in a large reactions dataset and understand the possible different types of reactions in the absence of any other information about the reactions. Note that the projected hypergraph is a dyadic, weighted directed graph obtained by using a role-interaction kernel that decomposes a hyperedge into a set of weighted directed edges. We perform community detection on such projected hypergraph. The alternative is to perform clustering directly using the hypergraph representation. However, given the scale of the hypergraph network of organic chemistry, the current clustering methods are computationally prohibitive.

\begin{figure*}[h]
    \centering
    \subfigure[][\footnotesize{Size distribution of identified communities}]{{\includegraphics[width=0.4\textwidth]{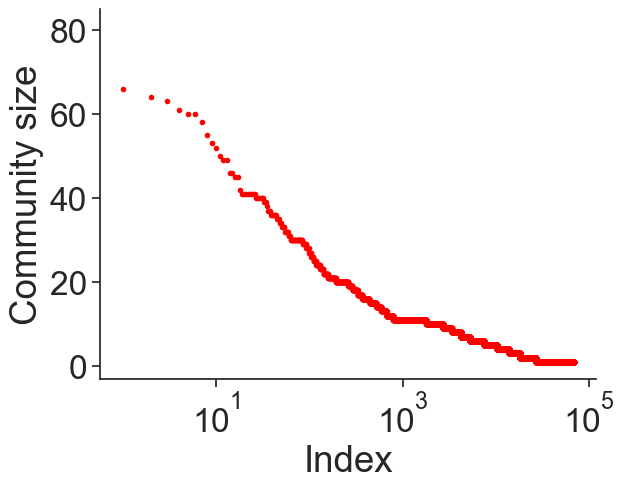}}}\quad
    \subfigure[][\footnotesize{Top-8 largest communities with nodes in each community in a different color}]{{ \includegraphics[width=0.4\textwidth]{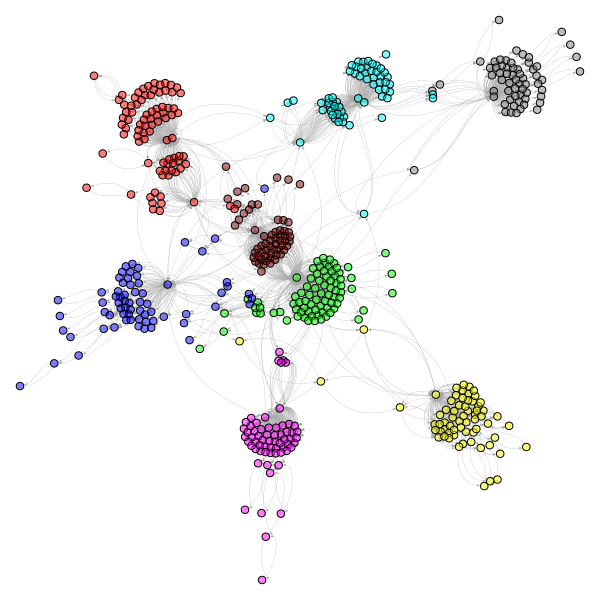}}}\\
    \subfigure[][\footnotesize{Top-100 largest communities indicating showing clear regions of high and low densities along with an island community disconnected from the rest of the network}]{{ \includegraphics[width=0.85\textwidth]{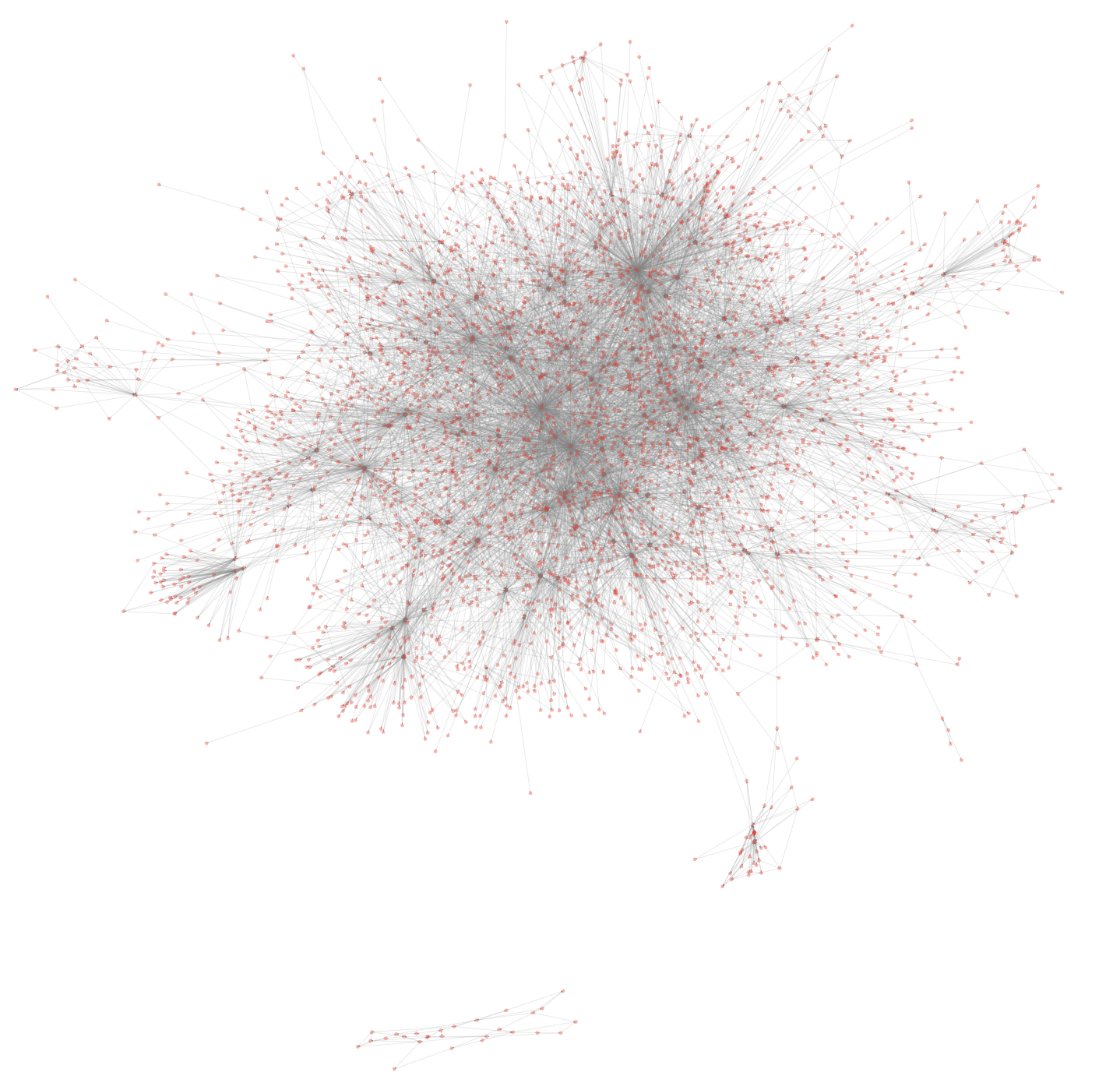}}}\\
    \caption{Community detection results on the weighted projected directed graph obtained using role-interaction kernel R2}
    \label{fig:communities_top12}
\end{figure*}

For the entire network, the Leiden algorithm identifies nearly $65,000$ communities with a size-distribution as shown in Figure \ref{fig:communities_top12}(a); the top-8 largest communities are shown in Figure \ref{fig:communities_top12}(b) with different color for each identified community, and the top-100 communities are show in Figure \ref{fig:communities_top12}(c). 

We observe from Figure \ref{fig:communities_top12}(a) that most of the communities are really small in size consisting of less than 10 reactions, whereas there are around 8 biggest communities containing over 60 reactions in each of them, as shown in Figure \ref{fig:communities_top12}(b). The close-knit nature of these communities point towards a possible segregation of different types of reactions just based on their connectivity patterns and the nodes (molecules) that take part in those reactions. This is the idea that we utilize to perform reaction type classification in the next section. Finally, the top-100 communities visualized in Figure \ref{fig:communities_top12} shows clear regions of high density with highly connected and localized clusters, and regions of low density further apart from the biggest clusters. In addition, it is also observed that there is a cluster that is completely separated from all the other communities and is therefore an \textit{island} community. The existence of core-periphery regions in the reaction network was also shown in \cite{bishop2006core} but the analysis was not based on graph clustering but on identifying strongly connected components in the network by representing reactions using a single directed edge from the heaviest reactant to the heaviest product in each reaction. In the community detection algorithm that we work with, we take into account the directionality as well as the weights of the edges, making it more flexible and the results more generalizable.

\section{Application in reaction class prediction problem} \label{sec:reaction-classif-results}
In the foregoing sections, we have shown how the hypergraph representation could be used to uncover hidden insights contained in large reactions datasets and study their time-evolution through network-theoretic properties. In this section, we demonstrate the usefulness of the hypergraph representation in capturing the context of reactions and thereby their reaction type or class. We, therefore, use the hypergraph representation in the reaction-type classification problem where the objective is to estimate the reaction class from a given set of reactants and products. This problem has practical applications in retrosynthetic planning where several different routes could be eliminated just by knowing the possible reaction types. The other problems where such a problem would find significance is the reaction feasibility estimation problem where the objective is to estimate the feasibility of a reaction given the possible participating molecules. Other studies that have proposed data-driven frameworks for reaction classification problems are \cite{probst2022reaction, baylon2019enhancing, schneider2015development}.

\subsection{Dataset description}
For this problem, since we require reaction class information for reactions, we use a subset of the USPTO reactions dataset that is typically used for retrosynthesis problem, containing about 50K reactions annotated with their corresponding reaction class from 10 possible classes. We generate the equivalent reaction hypergraph network for this dataset and work with the largest connected component in the hypergraph since the hyperedge (or reaction) embedding framework that we use to represent reactions subsequently in the classification framework is dependent on the connectivity and neighborhood contextual information. The distribution of reactions across different reaction classes in the sub-hypergraph is shown in Table \ref{tab:distributionreactionclass} below.

\begin{table}[h]
\centering
\caption{Distribution of reactions across different reaction classes in the largest connected subcomponent}
\label{tab:distributionreactionclass}
\resizebox{0.4\textwidth}{!}{%
\begin{tabular}{@{}ccc@{}}
\toprule
Rxn class & Rxn name                               & Num rxns \\ \midrule
1         & Heteroatom alkylation and arylation    & 11,526   \\
2         & Acylation and related processes        & 8,488    \\
3         & C--C bond formation                    & 3,909    \\
4         & Heterocycle formation                  & 588      \\
5         & Protections                            & 646      \\
6         & Deprotections                          & 760      \\
7         & Reductions                             & 459      \\
8         & Oxidations                             & 305      \\
9         & Functional group interconversion (FGI) & 1,168    \\
10        & Functional group addition (FGA)        & 196      \\ \bottomrule
\end{tabular}%
}
\end{table}

\subsection{Reaction embeddings using random hyperwalks}
To perform reaction classification by training a data-driven classifier, we need numeric representations for reactions that are generated from their hypergraph representations and would, therefore, be used as features to train a classifier. We generate hyperedge (or reaction) embeddings by adapting the deep hyperedges framework \cite{payne2019deep} and modifying it to incorporate the contextual information contained in chemical reactions, as explained in the pseudocode provide in Algorithm \ref{alg:random_hyperwalk}. 

The hyperedge embeddings are generated by performing random hyperwalks that capture the co-member information in each vertex by traversing hyperedges in the hypergraph network of chemical reactions. For each hyperedge, the hyperwalk starts at a randomly selected reactant node that is part of the current hyperedge, and either hops to a node in the adjacent hyperedge or stays in the current hyperedge to select another reactant node in the current hyperedge. This is repeated until the desired length of the hyperwalk is achieved. The adjacent hyperedge traversal is done only with respect to reactants since this would -- first, differentiate reactants from products, and second, mimic chemistry more realistically where only those reactions are accessible where either the current reactants participate as reactants or the product of the current reaction participates as reactant. Such a random hyperwalk would closely mimic a chemist performing experiments randomly.

Formally, we start at a node $v_m$ selected at random with the annotation `reactant' in a hyperedge $e_i$. The probability of traversing an adjacent hyperedge is inversely proportional to the cardinality of the current vertex; i.e. $p=min(\frac{\alpha}{\mid v_m \mid}+\beta, 1)$ where $\alpha$ and $\beta$ are tunable hyperparameters and $\mid v_m \mid$ is the cardinality of the vertex $v_m$. As in a random walk, if $p$ is less than a randomly generated number, the traversal is performed to an adjacent hyperedge; otherwise the current hyperedge is added to the random walk and the next is chosen randomly from the adjacent hyperedges  of the current vertex $v_m$. For each hyperedge $e_i$, we construct $50$ random walks of length $50$ each. Examples of such hyperedge random walks on the four example reactions in Equation \ref{eq:example_reactions} is shown in Table \ref{tab:hyperwalks_example}. The hyperwalks are then embedded into dense vectors of dimension $R^{256}$ using skip-gram approach for generating embeddings \cite{mikolov2013distributed}. At the end of the hyperedge embedding exercise, we would have a 256 dimensional vector for each hyperedge in the network.

\begin{figure*}[h]
    \centering
    \includegraphics[width=0.9\textwidth]{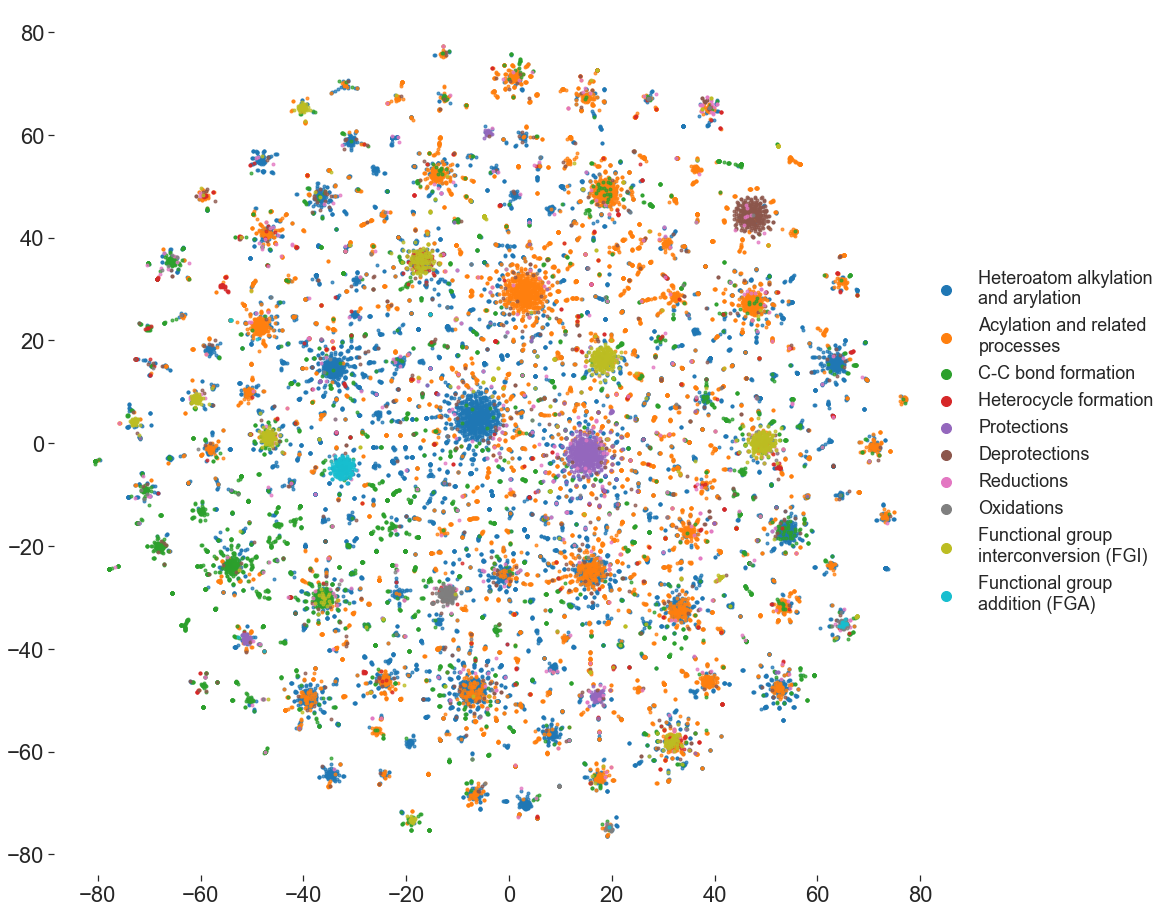}
    \caption{A 2D t-SNE projection of the 256-dimensional hyperedge embeddings}
    \label{fig:hyperedge_emb}
\end{figure*}

The pseudocode for the hyperwalk generating algorithm is presented in Algorithm \ref{alg:random_hyperwalk} and example hyperwalks using the example set of four reactions in Equation \ref{eq:example_reactions} is shown in Table \ref{tab:hyperwalks_example}. A 2D visualization of the resulting 256-dimensional hyperedge embeddings on the entire dataset of reactions is visualized in Figure \ref{fig:hyperedge_emb}.
\begin{algorithm}[H]
  \caption{Pseudocode for generating random hyperwalks for each hyperedge in the hypergraph}
  \label{alg:random_hyperwalk}
  \tiny
    \SetKwInOut{Input}{Input}
    \SetKwInOut{Initialize}{Initialize}
    \SetKwInOut{Output}{Output}
    
    \Input{\newline 
      walkLength:length of each hyperwalk \newline
      hyperEdges: set of hyperedges in the hypergraph \newline
      vertexMemberships: membership dictionary for each vertex indexed by vertex id and vertex role (product, reactant) \newline
      $\alpha$ and $\beta$: probability distribution parameters}
     \Initialize{\newline
     walks\_all = [ ]        \tcp*{stores all the hyperwalks generated}
     }
     \For{hyperedge\_id in hyperEdges}
     {curr\_walk = [ ]       \tcp*{stores hyperwalk for the current hyperedge}
     hyperEdge = hyperEdges[hyperedge\_id]
     \blockline
     curr\_vertex = randomly chosen `reactant' vertex in hyperEdge \tcp*{hyperwalk always starts from `reactant' nodes}
     curr\_hyperEdge = hyperEdge \newline
     \While{ len(hyperWalk) $<$ walkLength}
     {
      proba = $\alpha$/len(vertexMemberships[curr\_vertex][`reactant']+\\vertexMemberships[curr\_vertex][`product'])+$\beta$ \newline
     \If{random.random() $<$ \texttt{proba}}
     {
     adjacent\_vertices = curr\_hyperEdge[`reactant']+curr\_hyperEdge[`product']  \tcp*{switch to one of the adjacent vertices in current hyperedge}
     curr\_vertex = random.choice(adjacent\_vertices)     
     }
     curr\_walk.append(curr\_hyperedge) \blockline
     adjacent\_hyperedges = vertexMemberships[curr\_vertex][`reactant']  \tcp*{adjacent hyperedges defined with respect to reactant roles}
     curr\_hyperedge = random.choice(adjacent\_hyperedges)   \tcp*{randomly choose from one of the adjacent hyperedges}
     }
     walks\_all.append(walk\_hyperedge)
     }
    \Output{walks\_all}
\end{algorithm}
\begin{table}[H]
\centering
\tiny
\caption{Two example hyperwalks generated for each reaction (hyepredge) in the example set of reactions. For each walk, $v_i \overset{e_k}{\longrightarrow}v_j$ represents a walk along hyperedge $e_k$ via nodes $v_i$ and $v_j$. The hyperwalks for each hyperedge are the sequential collection of such $e_k$'s starting at that hyperedge.}
\label{tab:hyperwalks_example}
\begin{tabular}{@{}cc@{}}  
\toprule
$\mathbf{H_{id}}$ & \textbf{Hyperwalk} \\ \midrule
\multirow{2}{*}{0}    &       $  B, 0 \overset{1}{\longrightarrow} C \overset{2}{\longrightarrow}     C \overset{2}{\longrightarrow}  C \overset{3}{\longrightarrow}       D \overset{3}{\longrightarrow}  A \overset{0}{\longrightarrow} A \overset{3}{\longrightarrow} A \overset{3}{\longrightarrow}  A \overset{3}{\longrightarrow} A    $               \\
                      &      $  B,0 \overset{1}{\longrightarrow} B \overset{1}{\longrightarrow}     C \overset{2}{\longrightarrow}  C \overset{2}{\longrightarrow}        C \overset{3}{\longrightarrow}  C \overset{3}{\longrightarrow} C \overset{1}{\longrightarrow} C \overset{2}{\longrightarrow}  C \overset{1}{\longrightarrow} C    $                \\ \midrule
\multirow{2}{*}{1}    &   $  E,1 \overset{2}{\longrightarrow} E \overset{1}{\longrightarrow}     C \overset{2}{\longrightarrow}  C \overset{1}{\longrightarrow}        C \overset{2}{\longrightarrow} D \overset{3}{\longrightarrow} D \overset{3}{\longrightarrow} D \overset{2}{\longrightarrow}  E \overset{1}{\longrightarrow} E   $    \\
                      &  $  B,1 \overset{1}{\longrightarrow} E \overset{3}{\longrightarrow}    E \overset{2}{\longrightarrow}  E \overset{1}{\longrightarrow}        E \overset{1}{\longrightarrow}  E \overset{2}{\longrightarrow} D \overset{3}{\longrightarrow} A \overset{3}{\longrightarrow}  A \overset{0}{\longrightarrow} A    $                   \\ \midrule
\multirow{2}{*}{2}    &   $  E,2 \overset{1}{\longrightarrow} B \overset{1}{\longrightarrow}     B \overset{0}{\longrightarrow}  B \overset{0}{\longrightarrow}        A \overset{0}{\longrightarrow} A \overset{0}{\longrightarrow} A \overset{3}{\longrightarrow} D \overset{2}{\longrightarrow}  D \overset{2}{\longrightarrow} D    $    \\
                      &  $  C,2 \overset{2}{\longrightarrow} C \overset{1}{\longrightarrow}    C \overset{3}{\longrightarrow}  C \overset{3}{\longrightarrow}        C \overset{2}{\longrightarrow}  E \overset{1}{\longrightarrow} E \overset{2}{\longrightarrow} C \overset{2}{\longrightarrow}  C \overset{3}{\longrightarrow} A    $                   \\  \midrule
\multirow{2}{*}{3}    &   $  A,3 \overset{3}{\longrightarrow} A \overset{0}{\longrightarrow}     A \overset{0}{\longrightarrow}  A \overset{0}{\longrightarrow}        B \overset{0}{\longrightarrow} B \overset{1}{\longrightarrow} B \overset{0}{\longrightarrow} A \overset{3}{\longrightarrow}  D \overset{3}{\longrightarrow} D    $    \\
                      &  $  C,3 \overset{2}{\longrightarrow} C \overset{2}{\longrightarrow}    E \overset{2}{\longrightarrow}  E \overset{3}{\longrightarrow}        C \overset{1}{\longrightarrow}  C \overset{3}{\longrightarrow} C \overset{1}{\longrightarrow} B \overset{0}{\longrightarrow}  A \overset{0}{\longrightarrow} A    $                   \\ \bottomrule
\end{tabular}%
\end{table}

\subsection{Reaction class prediction results}
To predict the reaction classes, we train a one-vs-rest classifier based on support vector machines (SVM) that learns a multi-class classification decision boundary. We used a randomized cross validation search strategy to perform hyperparameter tuning of the SVM model with a radial basis function. A detailed description of the SVM model and the mathematical framework that underlies it is provided in \cite{mann2022hybrid}.

The precision and recall metrics for each of the reaction classes computed using the test-set containing unseen reactions at the training stage are shown in Figure \ref{fig:reaction_classification_performance} below.

\begin{figure*}[h]
    \centering
    \subfigure[][]{{\includegraphics[width=0.57\linewidth]{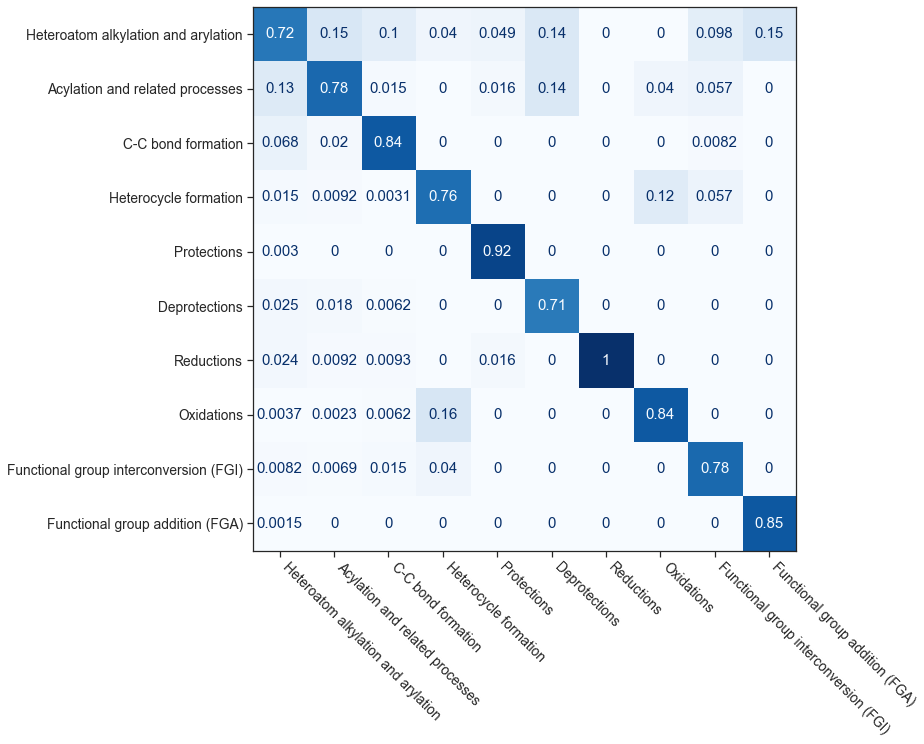}}}
    \subfigure[][]{{\includegraphics[width=0.423\linewidth]{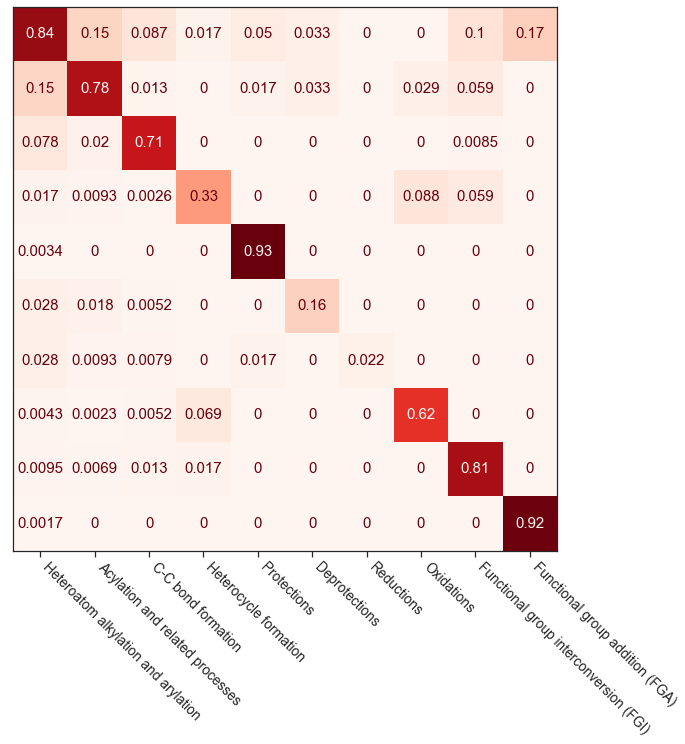}}}
    \caption{Performance metrics for the multi-class reaction classification on the test-set (a) Precision (b) Recall}
    \label{fig:reaction_classification_performance}
\end{figure*}

From the results above, we observe that the trained model accurately predicts the reaction class for most of the reaction classes except for the reaction classes -- reductions, deprotections, and heterocycle formation. The precision metrics across all 10 reaction classes shown in Figure \ref{fig:reaction_classification_performance}(a) highlights the model's high precision in identifying the correction reaction class. However, since the recall shown in Figure \ref{fig:reaction_classification_performance}(b) is lower for the three underperforming classes, there could be overlapping reaction classes in the feature (embedding) space. This is indeed observed for these classes in 2D visualization of the learned embeddings in Figure \ref{fig:hyperedge_emb}. The separation between various reaction classes could be addressed in future by also incorporating additional molecular descriptors that, in combination with the connectivity-specific embeddings, would more accurately distinguish the different reaction types. Nevertheless, the embeddings generated just based on the reactions and node-connectivity information in the hypergraph representation seems to have separated a majority of the reaction types into distinct clusters, consequently resulting in the model learning to predict them accurately. This again highlights the ability of hypergraphs to capture reaction context accurately.

\section{Conclusions and future work} \label{sec:conclusions}

Network theory offers natural tools and techniques for understanding the growth of chemistry over time by representing reactions as time-evolving real-world networks. Though most of the work in this area has been done using a dyadic graph representation, a hypergraph representation with hyperedges between nodes for representing reactions is a more natural, intuitive, and flexible representation that allows for the incorporation of additional reaction context.

We have shown that the hypergraph representation is more flexible, allows for incorporation of reaction-specific node context, and facilitates one-to-one correspondence of network properties with chemistry. We have computed detailed network statistics of the resulting hypergraph network of organic chemistry and studied the time evolution of these properties. As with several previous studies, we observed that the network exhibits a scale-free behavior with preferential attachment of nodes, has small average path length indicative of small-world nature, and shows assortative mixing with respect to certain node types. For all the network statistics presented, namely, degree distributions, average path length, assortativity or degree correlations, PageRank analysis, and community detection, we have correlated them with chemistry inferences that could be drawn from such analysis. In addition, we discovered that the network exhibits the phenomenon of initial attractiveness and network densification as chemistry evolves over time time.

To demonstrate the AI-applications of the hypergraph representation of chemical reactions, we performed reaction classification using embeddings generated from chemistry-informed random walks on hyperedges. The embeddings resulted in well-separated clusters for different reaction classes and consequently accurate reaction classification results. In future, we plan to extend this study on diverse (and possibly bigger) datasets across various subdomains, incorporate additional molecular descriptors for generating hyperedge embeddings for reaction classification, utilize the results in a retrosynthestic planning framework, and perform hyperedge prediction to discover new reactions.

\section*{Conflicts of interest}
There are no conflicts of interest to declare.

\section*{Author contributions}
\textbf{Vipul Mann}: Conceptualization, Formal Analysis, Writing - Original draft, Writing - review \& editing Methodology, Software; \textbf{Venkat Venkatasubramanian}: Conceptualization, Writing - review \& editing, Supervision, Funding acquisition

\section*{Acknowledgements}
This work was supported by the National Science Foundation (NSF) under Grant No. 2132142 and carried out at Columbia University.

\bibliography{ref.bib}

\end{document}